**Exploring genetic variation in the tomato (*Solanum* section *Lycopersicon*) clade by whole-genome sequencing**

The **100** Tomato Genome Sequencing Consortium

Saulo A. Aflitos[1,3], Elio Schijlen[1], Richard Finkers[2], Sandra Smit[3], Jun Wang[4], Gengyun Zhang[5], Ning Li[6], Likai Mao[7], Hans de Jong[8], Freek Bakker[9], Barbara Gravendeel[10], Timo Breit[11], Rob Dirks[12], Henk Huits[13], Darush Struss[14], Ruth Wagner[15], Hans van Leeuwen[16], Roeland van Ham[17], Laia Fito[18], Laëtitia Guigner[19], Myrna Sevilla[20], Philippe Ellul[21], Eric W. Ganko[22], Arvind Kapur[23], Emmanuel Reclus[24], Bernard de Geus[25], Henri van de Geest[1], Bas te Lintel Hekkert[1], Jan C. Van Haarst[1], Lars Smits[1], Andries Koops[1], Gabino Sanchez Perez[1], Dick de Ridder[3], Sjaak van Heusden[2], Richard Visser[2], Zhiwu Quan[5], Jiumeng Min[7], Li Liao[7], Xiaoli Wang[7], Guangbiao Wang[7], Zhen Yue[7], Xinhua Yang[7], Na Xu[7], Eric Schranz[9], Eric F. Smets[10], Rutger A. Vos[10], Han Rauwerda[11], Remco Ursem[12], Cees Schuit[13], Mike Kerns[15], Jan van den Berg[16], Wim H. Vriezen[16], Antoine Janssen[17], Torben Jahrman[18], Frederic Moquet[19], Julien Bonnet[22] and Sander A. Peters[1,*].

[1]Plant Research International, Business Unit of Bioscience, cluster Applied Bioinformatics, Plant Research International, Droevendaalsesteeg 1, 6708 PB Wageningen, The Netherlands. [2]Laboratory of Plant Breeding, Wageningen University and Research Centre, Droevendaalsesteeg 1, 6708 PB Wageningen, The Netherlands. [3]Bioinformatics, Department of Plant Sciences, Wageningen University and Research Centre, Droevendaalsesteeg 1, 6708 PB Wageningen, The Netherlands. [4]BGI-Shenzhen, Shenzhen, China. [5]State Key Laboratory of Agricultural Genomics, BGI-SZ. [6]BGI-Europe. [7]BGI-Shenzhen, Shenzhen 518083, China. [8]Laboratory of Genetics, Wageningen University and Research Centre, Droevendaalsesteeg 1,

6708 PB Wageningen, The Netherlands. [9]Biosystematics Group, Wageningen University, Droevendaalsesteeg 1, 6708 PB Wageningen, The Netherlands. [10]NBC, Darwinweg 2, 2333 CR Leiden, The Netherlands. [11]RNA Biology & Applied Bioinformatics, Swammerdam Institute for Life Sciences, Faculty of Science, University of Amsterdam, Science Park 904, 1098 XH Amsterdam, The Netherlands. [12]Rijk Zwaan Breeding B.V. department of R&D Bioinformatics, PO Box 40, 2678 ZG, De Lier, The Netherlands. [13]Bejo Zaden B.V., Research Centre, Trambaan 2A, 1749 CZ, Warmenhuizen, The Netherlands. [14]East West Seed, Hortigenetics Research, 7 Moo 9 Maefack Mai , Chiang Mai-Praw Rd., Sansai, Chiang Mai, 50290, Thailand. [15]Monsanto Holland B.V., Leeuwenhoekseweg 52, 2660 BB Bergschenhoek, The Netherlands. [16]Nunhems Netherlands B.V., PO Box 4005, 6080 AA Haelen, The Netherlands. [17]Keygene N.V. PO Box 216, 6700 AE Wageningen, The Netherlands. [18]Semillas Fito, Cetro de Biotecnologia, Riera d'Agell 11, 08349 Cabrera de Mar (BCN), Spain. [19]Gautier Semences, Route d'Avignon, 13630 Eyragues, France. [20]BHN Research PO Box 3267, Immokalee, FL 34143, USA. [21]CGIAR, Avenue Agropolis, F-34394 Montpellier Cedex, France. [22]Syngenta Biotechnology, Inc. 3054 East Cornwallis Road, PO Box 12257, Research Triangle Park, NC 27709-2257, USA. [23]Rasi Seeds (P) Ltd. 273, Kamarajanar Road, Attur-636 102, Salem District, Tamilnadu, India. [24]ADNid Cap Alpha, Avenue de l'Europe, 34830, Clapiers, France. [25]Stichting Technologisch Top Instituut Groene Genetica, Vossenburchkade 68A, 2805 PC Gouda, The Netherlands.

**Running title:** The 100 tomato genome project

*corresponding author; Sander A. Peters; tel: +31-317-481123; fax: +31-317-418094; e-mail: sander.peters@wur.nl

**Keywords**: *Solanum* genetic diversity, comparative sequence analysis, phylogenomics, chromosome rearrangement, chromosome evolution, *Solanum lycopersicon, Solanum pennellii, Solanum arcanum, Solanum habrochaites.*

**Word count**: total 7,482 words (excl. references, affiliates, supplementary data); summary 269 words; introduction 1,048 words; results (excl. legends) 3,368 words; discussion 1,232 words;

experimental procedures 1,209 words; acknowledgements 41 words; legends 179 words; references 1,577 words; authors list 136 words; authors affiliations 293 words.

**SUMMARY**

Genetic variation in the tomato clade was explored by sequencing a selection of 84 tomato accessions and related wild species representative for the *Lycopersicon*, *Arcanum*, *Eriopersicon*, and *Neolycopersicon* groups. We present a reconstruction of three new reference genomes in support of our comparative genome analyses. Sequence diversity in commercial breeding lines appears extremely low, indicating the dramatic genetic erosion of crop tomatoes. This is reflected by the SNP count in wild species which can exceed 10 million i.e. 20 fold higher than in crop accessions. Comparative sequence alignment reveals group, species, and accession specific polymorphisms, which explain characteristic fruit traits and growth habits in tomato accessions. Using gene models from the annotated Heinz reference genome, we observe a bias in dN/dS ratio in fruit and growth diversification genes compared to a random set of genes, which probably is the result of a positive selection. We detected highly divergent segments in wild *S. lycopersicum species*, and footprints of introgressions in crop accessions originating from a common donor accession. Phylogenetic relationships of fruit diversification and growth specific genes from crop accessions show incomplete resolution and are dependent on the introgression donor. In contrast, whole genome SNP information has sufficient power to resolve the phylogenetic placement of each accession in the four main groups in the *Lycopersicon* clade using Maximum Likelihood analyses. Phylogenetic relationships appear correlated with habitat and mating type and point to the occurrence of geographical races within these groups and thus are of practical importance for introgressive hybridization breeding. Our study illustrates the need for multiple reference genomes in support of tomato comparative genomics and *Solanum* genome evolution studies.

## INTRODUCTION

The *Solanaceae* or Nightshade family consists of more than 3000 species covering a very large diversity in terms of habit, habitat and morphology. Its species occur worldwide growing as large forest trees in wet rain forests to annual herbs in deserts (Knapp, 2002). *Solanum* is the largest genus in the family and includes tomato (*Solanum lycopersicum*) and various other species of economic importance. Tomato breeding during the past decades focused on higher productivity and adaption to different growing systems. Its economic success is reflected by the fact that, on a global scale, tomato is one of the most important vegetable crops with a worldwide production of 159 million tons covering some 4,700,000 ha (www.fao.org). Yet, domestication for tomato has been clearly distinct from the species divergence by natural selection as a consequence of selecting for a limited set of traits, including attractive red fruit color and size. As a result its genetic basis was seriously narrowed, known as the 'domestication syndrome' (Hammer, 1985; Doebley *et al.*, 2006; Bai and Lindhout, 2007; Bauchet *et al.*, 2012). In more recent times tomato was adapted to different growing systems involving a small number of traits, including self-pruning, plant height, earliness, fruit morphology and non-red fruit color (Bauchet *et al.*, 2012). The relative small genetic variation became apparent in the face of rapidly changing environmental conditions, competing claims for arable lands and new consumer requests. These challenges push tomato breeding efforts towards better biotic and abiotic stress tolerance, higher productivity and increased sensory and nutritional value. However, the reduced genetic variation that resulted from extensive inbreeding has decelerated tomato crop improvement. To enlarge the genetic basis, breeders now focus on introgression of desirable genes from the wild relatives into their elite cultivars, which so far has been quite limited (Singh, 2007; Bai and Lindhout, 2007).

The first step of introgressive hybridization includes crosses of the cultivated tomato with its heirlooms, wild relatives or more distant species of the tomato clade. Introgression breeding is practicable as cultivated tomato and related wild species are intracrossable, and most of the wild species are also intercrossable (Rick, 1979; 1986; Spooner *et al.*, 2005) despite the fact that diverse mating systems have evolved varying from allogamous self-incompatible (SI) and facultative allogamous to autogamous self-compatible (SC). Especially at the geographic margins of the distributions, interspecies changes in incompatibility systems that promote inbreeding over outcrossing have been documented (Peralta *et al.*, 2008; Grandillo *et al.*, 2011). Species boundaries and genetic diversity have been extensively studied in tomato using a wide range of molecular data (reviewed in Peralta *et al.*, 2008 and Grandillo *et al.*, 2011). For example, RFLP analysis showed that genetic diversity for SI species far exceeds that of SC species, estimated at 75% vs. 7% (Miller and Tanksley, 1990). Furthermore, 'within-accession' genetic variation was estimated at 10% of the 'between-accession' variation, in contrast to the genetic variation of the modern cultivars estimated at less than 5%. This further illustrates the dramatic erosion of genetic diversity in cultivated tomato crops.

Selection of crossing parents for interspecific hybridization requires insight in phylogenetic relationships for the tomato clade, but their tree based on morphological and molecular data has not been undisputed. Four informal species groups were proposed for the tomato clade, *Lycopersicon*, *Arcanum*, *Eriopersicon* and *Neolycopersicon* (Peralta *et al.*, 2008), which are supposed to have evolved from a most recent common ancestor approximately 7 million years ago (Nesbitt and Tanksley, 2002; Spooner *et al.*, 2005; Moyle, 2008, Peralta *et al.*, 2008). In spite of these studies, evolutionary relationships between the 13 species in the *Lycopersicon* clade are not fully resolved, considering the dichotomy between *Solanum pennellii* and *Solanum habrochaites* (Spooner *et al.*, 2005; Peralta *et al.*, 2005; 2008). The evolutionary history of *Solanum* genomes has also been investigated from the perspective of chromosome organization. The study of Szinay *et al.* (2012) on cross-species BAC FISH painting of *Solanum* species revealed few large rearrangements in the short arm euchromatin of chromosomes 6, 7 and 12, whereas Anderson *et al.* (2010) demonstrated pairing loops, multivalents and kinetochore shifts in synaptonemal complex spreads of hybrids between different members of the tomato clade, hinting at paracentric and pericentric inversions and translocations between the homeologous chromosomes. Furthermore, comparative genomics point to a *Solanum* genome landscape in which chromosome evolution for the majority of the 12 chromosomes has been far more dynamic than currently appreciated (Peters *et al.*, 2012). Collectively, these findings demonstrate that evolutionary relationships among the wild relatives still have to be considered provisional (Peralta *et al.*, 2008).

The availability of high-throughput sequencing technologies has provided unprecedented power to determine genome variation across entire clades, both at the structural and the genotype level. Initiatives such as the 1001 genomes project for *Arabidopsis thaliana*, the *Drosophila* sequence project, and the 1000 genomes project for human have been illustrative for the discovery of a vast amount of intraspecies specific polymorphic sequence features like InDels, repeats and SNPs for hundreds of genes (Weigel and Mott, 2009; Mackay *et al.*, 2012; The 1000 genomes project consortium, 2010), and have illustrated that there is no such thing as “the genome” for a particular species. Rather, the range of physiological and developmental traits appears to be reflected in the tremendous amount of sequence variants contributing to intraspecific variation. Considering the overwhelming interspecies genetic variability, tomato germplasm collections represent a gene pool with unprecedented possibilities to address new breeding demands imposed by climate change, world population increase, and consumer needs. Here we aim to reveal and study this genetic variation by genome-sequencing a selection of representative tomato accessions, which has become attainable with the recent development of the *S. lycopersicum* Heinz 1706 reference genome (The Tomato Genome Consortium, 2012). In addition to this reference genome for the *Lycopersicon* species, we present the construction of reference genomes of three other related species representing the *Arcanum*, *Eriopersicon* and *Neolycopersicon* group, respectively, providing an expanded resource for detailed comparative genomic studies in the near future. We also present results on robust/high confidence detection and identification of sequence polymorphisms, heterozygosity levels, introgressions, and assess the genetic diversity within the tomato clade from a phylogenetic and evolutionary perspective. This study provides an invaluable dataset for advanced omics studies on sequence trait relationships, the molecular mechanisms of tomato genome evolution as well as developing genotyping-by-sequencing breeding approaches.

## Results

### Selection of tomato accessions

We have selected 84 accessions of the *Solanum* clade section *Lycopersicum* for shallow (36 fold coverage) whole genome sequencing. A first set of 54 accessions consists of tomato landraces and heirloom cultivars of *S. lycopersicum* and *S. lycopersicum* var. *cerasiforme* which have been selected from the EU-SOL tomato core collection (https://www.eu-sol.wur.nl). The second set of 30 accessions comprises wild relatives of tomato representing the full range of expected genetic variation around *S. lycopersicum*. Their selection was based on previous usage in genetic research and previous utilization of quality or (a)biotic stress traits (reviewed in Grandillo *et al.*, 2011). We also chose *S. arcanum* LA2157, *S. habrochaites* LYC4 and *S. pennellii* LA0716 for *de novo* sequencing and whole genome reconstruction, aiming to have a reference genome available for each of the four main phylogenetic groups in the tomato clade. An important selection criterion was the self-compatibility of these accessions allowing inbreeding for several generations to minimize heterozygosity, and so reduce *de novo* genome assembly problems. A complete list of the selected accessions used in this study can be found in table S1.

### *De novo* assembly of three wild tomato relatives and Heinz

Comparisons of molecular data have indicated relatively low DNA sequence diversity between genetically related species within the phylogenetic groups of the tomato clade (Miller and Tanksley, 1990). Furthermore, preliminary analysis indicated that SNP frequencies for *S.*

*pimpinellifolium* and *S. pennellii* compared to *S. lycopersicum* were 1% and 10% respectively. Considering that *S. pimpinellifolium* and *S. pennellii* phylogenetically are among the closest and most distantly related species to tomato respectively, we assumed the same range for the other species. Our strategy to determine the proportion of polymorphic loci across the entire *Lycopersicon* clade was therefore targeted at *de novo* sequencing and assembly of three new reference genomes, followed by shallow sequencing of the bulk of the accessions and subsequently mapping them to a reference genome. For reference genome reconstruction we relied on massive parallel sequencing using Illumina HiSeq 2000 and 454FLX technology. Two paired-end libraries with insert sizes of 170 bp and 500 bp and one mate pair library of 2 kbp were sequenced using Illumina at 25, 25, and 30 fold coverage (assuming a genome size of 950 Mbp) respectively, and had at least 80% of the bases with Q-value above 30 (error rate <= 1/1,000). For the 454FLX sequencing, large insert size libraries of 8 kbp and 20kbp were created each at 0.6 coverage. *S. pennellii* LA0716 had an additional 8 kbp Illumina mate pair library at 0.4 fold. We discarded unpaired reads resulting in 205 fold coverage. For *de novo* assembly we aimed at maximizing short-range contiguity, long-range connectivity, completeness and quality by following the strategy as outlined by Gnerre *et al.* (2010). Our assembly statistics show a total contig length for *S. arcanum*, *S. habrochaites* and *S. pennellii* reaching a plateau of approximately 760 Mb (figure 1, table S2). The unique portion is comparable in size in these genomes, which is consistent with widespread research including comparative mapping studies revealing a high level of synteny among the species of the *Solanaceae* (Paterson *et al.*, 2000). However, previous estimates on DNA content and flow cytometry analyses suggest a considerable variation in total genome size among species in the tomato clade (reviewed by Grandillo *et al.*, 2011). For example, the DNA content of cultivated tomato varies from 1.87 to 2.07 pg/2C indicating to a genome size of approximately 950 Mbp, whereas that of *S. pennellii* is substantially larger and corresponds to a DNA content of 2.47 to 2.77 pg/2C corresponding to 1,200 Mbp. Furthermore, we assume that most of the estimated 35,000 genes reside on the ~220-250 Mb of DNA in the euchromatic regions (http://www.rbgkew.org.uk/cval/; Arumuganathan and Earle, 1991; Van der Hoeven *et al.*, 2002; The Tomato Genome Sequencing Consortium, 2012). The increased genome size of *S.*

*pennellii* is likely to a greater part explained by an expansion of the repetitive portion of the genome. Repeats are known to impede genome reconstruction resulting in a more fragmented assembly and a lower N50 contig size. This is consistent with the *S. pennellii* LA0716 assembly statistics (table S2). The re-assembled *de novo S. lycopersicum* cv Heinz reaches the assembly size plateau more slowly and also appears more fragmented than the published reference genome, which likely is due to the use of older sequencing platforms and the BAC-by-BAC sequencing strategy that was used for this species previously.

To determine the extent of sequence diversity, read pairs from *de novo* sequenced genomes were mapped to the *S. lycopersicum* cv Heinz 1706 v2.40 reference genome. The lowest number of unmapped reads (11%), which likely consists of low quality sequences and increases for *S. arcanum* (17%), *S. pennellii* (22%) and *S. habrochaites* (25%), respectively. Given the comparable sequence quality we assume an equal percentage of low quality reads for the *de novo* sequenced genomes, while sequence diversity, introgressions and genome expansion contribute to the remainder of the unmapped reads.

**Sequencing and mapping of the 84 accessions and wild species**

For the 84 accessions $2.9 \times 10^{12}$ base pairs were sequenced equaling to an average coverage of 36.7±2.3 (32.5±2.1 with Q >= 30) fold per accession. All individuals were mapped against *S. lycopersicum* cv Heinz 1706 v2.40 to assess the diversity in both crop and wild-species, resulting in 96.4%±0.88% and 52.9%±2.93% of the reads correctly mapping for crops and wild species, respectively (fig. 2). These numbers improved when reads from wild species were mapped against a new reference genome from a closer relative. For *S. arcanum*, S. *habrochaites*, and *S. pennellii* 72.87%±7.87%, 78.74%±15,63%, and 55.37%±9.29% of the reads correctly mapped against the *S. arcanum* LA2157, *S. habrochaites* LYC4, and *S. pennellii* LA0716 reference genome respectively. These results illustrate the large genetic erosion within the crop tomatoes and the large sequence diversity among the wild species. Moreover, it emphasizes the need for multiple reference genomes to support interpretations of genetic variation consequences among species in the tomato clade, which would otherwise be biased toward a single reference genome that is genetically more distantly related to the wild species.

## Whole genome sequence diversity

To further assess the sequence diversity in *Solanum* section *Lycopersicon* we quantified and classified the SNPs for each of the 84 accessions using read mappings against Heinz. The SNP counts for tomato cultivars are relatively low and gradually increase for *S. galapagense*, *S. cheesmaniae* and *S. pimpinellifolium* accessions. Specific members of the *Arcanum*, *Eriopersicon*, and *Neolycopersicon* groups SNP numbers sharply increase (figs. 3 and S1), which correlates with their more distant position in the phylogenetic tree in the tomato clade (Peralta *et al.*, 2008).

When compared to the Heinz annotated genome, in all accessions we consistently observed a significant higher SNP frequency in intergenic regions than in genic regions. Approximately, 89.47%±3.03% of the polymorphisms falls into intergenic regions, while 7.55%±2.19% maps to introns and 2.33%±0.68% maps to exons (fig. 4). Of the polymorphisms in exons, 55.17%±11.54% is synonymous while 44.83%±21.03% is non-synonymous (fig. S2).

The number of SNPs in wild species on average appears 20 times higher than in crop tomatoes. These results are consistent with the notion that crop tomato genomes are extensively genetically eroded compared to the large genetic diversity found among the wild species. A striking trend is the genome wide ratio between synonymous and non-synonymous SNPs (dN/dS). For crops, non-synonymous SNPs outnumber synonymous SNPs while the opposite is generally true for wild species (fig. S2). Although we currently have no clear explanation for the higher dN frequency in crop accessions, it might partly be the result of the artificial selection pressure imposed by breeding, maintaining only a relatively small number of SNPs under positive selection while allowing the fixation of many non-synonymous SNPs as has previously been reported for tomato by comparative transcriptomics (Koenig *et al.,* 2013).

A JBrowse (Skinner *et al.*, 2009; Westesson *et al.*, 2013) supported overview of the SNP and InDel variation in the 84 accessions can be accessed in the tomato 100+ variant browser that is publicly accessible via http://www.tomatogenome.net/VariantBrowser/.

## Heterozygosity and introgressions

For the *lycopersicum* accessions, highest heterozygosity levels were observed for beef type accessions *S. lycopersicum* EA03222 and EA01155 (Dana) as shown in figure 3. With respect to mating type, highest ratios were found for allogamous SI wild species, while facultative SC wild species display an intermediate heterozygosity ratio (fig. S3). On average autogamous SC species have a slightly lower heterozygosity level compared to facultative SC species, of which the autogamous SC wild species *S. neorickii* LA0735 has the lowest (fig. S3).

Surprisingly, some tomato accessions display considerable high SNP counts (figs. 3 and S1) which might be attributed to introgressions. Indeed, additional footprints for interspecific introgressions in 54 *lycopersicum* accessions compared to Heinz were found by testing the SNP distributions in 1 Mbp sized bins along each of the 12 chromosomes. Bins with SNP counts deviating significantly from the average (z-test $p < 0.05$) were considered as either introgressed in crop accessions or natural highly divergent in wild *S. lycopersicum* species after subtracting the number of SNPs found when mapping Heinz reads against itself and correcting for chromosome position and species specific effects by median polish. Introgressions appeared in 5.56% ± 7.98% of the 767 1Mbp bins with *S. lycopersicum* PC11029 having the smallest number (0.13%), *S. lycopersicum* LA0113 having the highest number (31.42%), and *S. lycopersicum* PI272654 with 0.91% of its bins marked as introgression. Cherry, giant and beef tomatoes have a higher number of introgressions among the crops, while wild *S. lycopersicum* species are even more divergent (fig. S4).

The specific chromosome locations of divergent SNP intervals are displayed in figure S4. Here, similar patterns of SNP concentrations can be observed between crop accessions which most likely are introgressions originating from the same donor accessions. In some cases the most likely source of introgressions could be deduced from the SNP identity and the phylogenetic distance inferred from the SNP alignment. Indeed, when plotting the chromosomal SNP distribution, we found a 2.2 Mb introgressed segment in the long arm of chromosome 6 roughly between Tomato-EXPEN2000 genetic markers C2_At4g10030 (44 cM) and TG365 (50 cM) for the accessions LA2838A (Alisa Craig), LA2706 (MoneyMaker), LA2463 (All Round) and CGN15820. Phylogenetic distance analysis reveals a 2.2Mb segment in the heirloom open pollinated tomato accession MoneyMaker is most closesly related to the wild species *S.*

*pimpinellifolium* LYC2798 (fig. S5). Interestingly, the Heinz ITAG 2.4 annotation of this segment points to several loci that have been implicated in hormone induced stress responses, fruit development, flavonoid phytonutrient production, and MAPK mediated production of reactive oxygen species involved in innate immunity to *Phytophtera infestance* induced late blight.

**Sequence diversity and phylogenetic relationships**

*SNPs in genes related to fruit and growth diversification*

To analyze diversity in specific genes and loci which underlie a phenotypic effect on Fruit Diversification and Plant Growth (FDPG), we determined the orthologs for *ovate* (Solyc02g085500), *fw2.2* (Solyc02g090740), *ls* (Solyc07g066250), *og/beta* (Solyc06g074240), *lcy1* (Solyc04g040190), *lfy* (Solyc03g118160), *rin* (Solyc05g012020), *sp* (Solyc06g074350), *fer* (Solyc06g051550), *style* (Solyc02g093580), *psy1* (Solyc03g031860), *lin5* (Solyc09g010080), and for locus *lc* (gb|JF284941). Based on the ITAG 2.4 annotation of the tomato reference genome (The Tomato Sequencing Consortium, 2012), the polymorphisms in the orthologs were classified as coding or non-coding, and as non-synonymous or synonymous (silent) SNPs to compare intra- and interspecies sequence diversity and SNP effects among the 84 accessions. FDPG genes in many cases underlie a phenotypic trait that is determined by a few SNPs or sometimes a single one (see below). While tomato breeding has primarily been directed toward selection of these traits, it is conceivable that a SNP determining a single trait went through a positive selection, whereas the bulk of the genes were subjected to a more relaxed selection. Non-synonymous SNP counts in FDPG genes from wild species are consistently higher than observed for a randomly selected set of genes. In contrast, synonymous SNP are consistently lower. Nevertheless, both counts are just within 1 standard deviation away from the average (fig. S6). Perhaps this observation reflects a higher selection pressure in wild species than in crop accessions against deleterious mutations in FDPG genes.

*Lycopersicon, Aracanum, Eriopersicon and Neolycopersicon* specific SNPs

Several characteristic SNPs were found distinctive for the *Lycopersicon*, *Arcanum*, *Eriopersicon* and *Neolycopersicon* section. For example, the red or orange to yellow fruited *Lycopersicon* group accessions have a GTC codon in the *og/beta* gene of tomato chromosome 6 for the $Val_{23}$ amino acid in the chromoplast specific lycopene beta cyclase, whereas the green-fruited *Arcanum, Eriopersicon and Neolycopersicon* species have a non-synonymous TTC (Phe) substitution. The *lcy1* gene on chromosome 4, which has GAG codon for the $Gln_{30}$ residue of lycopene beta cyclase 1, has been substituted in all accessions in the *Arcanum* group. In particular, *S. chmielewskii* accessions have a GTG (Val), while the *S. neorickii* and *S. arcanum* accessions have a CTG (Leu) codon.

*Species specific SNPs*

In addition to group specific polymorphisms, we also found species specific SNPs. For example, further downstream in the *og/beta* orthologous gene of *S. corneliomulleri* a GCT ($Ala_{437}$)>ACT (Thr) and a TTG ($Leu_{464}$)>TTT (Phe) SNP occur, whereas the AAA codon for amino acid $Lys_{277}$ in tomato is substituted into an ATA (Ile) for the *S. chmielewskii* accessions. In the *lfy* gene orthologs, a synonymous SNP TTA ($Leu_{25}$)>CTA is shared by the *huaylasense* accessions, whereas a CCA ($Pro_{122}$)>CAA (Gln) nucleotide substitution, is characteristic for *S. chmiemlewskii* accessions. In the *fer* gene *S. peruvianum* accessions have a CGATGA insertion (AspAsp) downstream and adjacent to ($Asp_{89}$). *S. arcanum* and *S. chilense* accessions share a GCC ($Ala_{107}$)>GCA synonymous SNP in the *ovate* gene ortholog, whereas we detected several intron SNPs that are specific for *S. neorickii* in the *sp* orthologous gene. Finally, in the *style* gene of *S. lycopersicum* accessions a TTT ($Phe_{80}$)>TTC substitution is characteristic for *S. chilense* accessions.

*Accession specific SNPs related to fruit traits*

We also observed accession specific polymorphisms related to specific fruit traits. Black Cherry has a single nucleotide deletion in the coding sequence of the chromosome 6 *B* gene (fig. S7). This specific deletion occurs in the *old-gold-crimson* ($og^c$) null allele (Ronen *et al.*, 2000). The resulting frame shift causes the loss of lycopene-*ß*-cyclase function underlying the accumulation of lycopene and dark red/purple appearance of tomato fruits, and thus is likely to be the cause for the characteristic dark red/purple flesh-coloured fruits of Black Cherry. Galina, Iidi and T1039 are yellow-skin cherry tomatoes, and have a single nucleotide deletion resulting in a frame shift causing a $Lys_{389}$>Ser substitution and a premature TGA stop codon directly downstream that would result in a truncated *psy1* protein lacking the terminal 23 amino acids (fig. S7). In this respect it is interesting to note that the $r^y$ mutant allele, which encodes a phytoene synthase lacking these terminal amino acids, underlies the yellow-coloured fruit skin phenotype in tomato mutants (Fray and Grierson, 1993).

Fruit shape and size in tomato is influenced by locule number. Two QTLs, *lc* and *fas* have major effects on these traits and can act synergistically leading to extreme high locule numbers (Cong *et al.*, 2008; Munos *et al.*, 2011). *Fas* is the major gene responsible for increasing locule numbers from 2 to more than 6, while *lc* has a weaker effect increasing locule numbers to 3 or 4. Two T>C and A>G SNPs are associated with the high locule number allele ($lc^h$), while an extreme high locule number caused by down regulation of a YABBY-like transcription factor is associated with a 6-8kb insertion in the first intron of the *fas* gene (Cong *et al.*, 2008). Sequence analysis revealed that all bilocule accessions have the low locule number allele ($lc^l$), while accessions with 3 to 4 locules (except Cal J TM VF and Dana) have the $lc^h$ allele. Pear-shaped tomato fruit is controlled by the quantitative trait locus *OVATE*. The allelic interactions at the ovate locus have been described as recessive but their expression depends on the genetic background (Ku *et al.*, 1999). Liu and co-workers (2002) showed that a GAA ($Glu_{279}$)>TAA non-sense mutation in the second exon causes an early stop codon and a premature translation termination resulting in a 75 amino acid truncated ovate protein (AAN17752) leads to pear-shaped fruit formation. All accessions with pear-shaped fruits have the premature stop codon, while the mutational effect is less pronounced in the ovate-fruited accession ‘Porter’ (fig. S8). Hereafter, we address sequence diversity in view of the intraspecies and interspecies phylogenetic relationships.

**Phylogenetic relationships**

Cladistics based on molecular data resulted in the clear grouping of species within the *Solanum* genus section *Lycopersicon* (Peralta *et al.*, 2008). However, at the species level, relationships are still unresolved. For example, while *S. pennellii* was placed in its own group *(Neolycopersicon)* as a sister to the rest of the section *Lycopersicon*, it nonetheless appeared as sister to *S. habrochaites* in the main trichotomy (Spooner *et al.,* 2005; Peralta *et al.*, 2008;

Grandillo *et al.*, 2011). Our SNP analysis indicates that many polymorphisms are distinct for *habrochaites* species, whereas *S. pennellii* LA0716 shares many SNPs with accessions of the *Arcanum* and *Eriopersicon* groups. This point to a complicated phylogenetic relationship for *S. pennellii* and *S. habrochaites*.

We applied the vast amount of multilocus molecular data to shed more light on the species and accession relationships in the tomato clade. First, we used a limited set of polymorphisms to assess the species boundaries and relationships within the tomatoes and wild relatives. The strict consensus tree for ten concatenated genes (fig. S9) revealed that all *S. habrochaites* species cluster into a monophyletic group, while *S. pennellii* LA0716 is sister to *S. habrochaites*. The *S. chilense* accessions also group together and cluster with *S. corneliomulleri* and *S. peruvianum* accessions, which are representatives of the former *S. peruvianum* 'southern group', and with accession LA2172. The green-fruited self-compatible (SC) *S. chmielewskii*, and *S. neorickii* species, which are representatives of the *Arcanum* group (Peralta *et al.*, 2008), are resolved into two monophyletic groups and cluster with two *S. arcanum* species into a larger clade. Furthermore, all red or orange-fruited SC species of the *Lycopersicon* group (*S. cheesmaniae*, *S. galapagense*, *S. lycopersicum*, *S, pimpinellifolium*) form a well-supported clade. In particular, the orange-fruited *S. cheesmaniae* and *S. galapagense* cluster into a subgroup illustrating the very close relationship between both species. These relations are in agreement with previously presented phylogenetic studies (reviewed by Grandillo *et al.*, 2011).

Next we excluded heterozygous SNPs from the analysis, as they are arbitrarily converted into a single nucleotide call for FASTA converted sequences thereby introducing noise and a possible bias in the data. SNPs in introns were also excluded as they are likely to be under less selective pressure than exon SNPs and probably carry less phylogenetic information and introduce more noise. Figure S10 shows the homozygous SNPs in the FDPG genes have sufficient power to resolve the phylogenetic placement consistent with the grouping at the sectional level as previously described (Peralta *et al.*, 2008). We noticed a slight increase in resolution when comparing the gene tree based on unfiltered and filtered SNPS respectively (figs. S9 and S10). Nevertheless, at the species level the placement of *lycopersicum*, *pimpinellifolium*, *galapagense* and *cheesmaniae* accessions appeared largely unresolved. In this analysis, *S. pennellii* is a sister species from the *Arcanum* group. We therefore also assessed the clustering using genome wide homozygous SNPs. The whole genome SNP cladogram in figure 3 shows a complete resolution into separate branches with high bootstrap values for each of the *Lycopersicon* accessions and wild species. Although phylogenetic relationships might be influenced by SNPs that arise from introgressions, the genome wide SNP information generates sufficient resolution power and enables the interspecies and intraspecies identification of all 84 individuals in monophyletic groups. Based on our phylogenetic analysis and SNP sequences we propose to type accession LYC2740 as an *S. lycopersicum* species instead of a *S. pimpinellifolium*. We also observed several *S. lycopersicum* accessions grouping with *S. pimpinellifolium*, *S. galapagense* and *S. cheesmaniae*. Those *S. lycopersicum* accessions likely are hybrids or carry substantial *S. pimpinellifoulium* introgressions. Additional analysis should be performed to substantiate this hypothesis. In addition, *S. pennellii* appears a sister species to *S. habrochaites* species group in the whole genome SNP tree, suggesting *S. pennellii* can be considered an intermediate species between *S. habrochaites* and *S. arcanum* which would coincide with its intermediate geographical distribution.

## Discussion

### *Multiple reference genomes and sequence diversity*

Our study has yielded a huge amount of precious data on sequence diversity in wild species of the tomato clade. The reads for *S. habrochaites* (78%), *S. Arcanum* (73%) and *S. pennellii* (53%) were mapped onto the corresponding species reference genome illustrating the large interspecies sequence variation in *the Lycopersicon* clade. We also demonstrated dramatic genetic erosion in cultivated tomatoes. As there is an increasing demand for broadening the genetic base of this crop we believe that our study provide pivotal information for future tomato breeding programs. The Heinz reference genome is not only partly representative for the genetic and structural information in the related wild species but it also emphasises the need to reconstruct additional reference genomes. The three *de novo* sequenced genomes presented here thus constitute a valuable additional resource to the currently available genomic tools in support of studies on evolution, domestication and genetic bases underlying important traits like disease resistance and abiotic stress tolerance.

Our sequencing and mapping strategy effectively supports the detection and identification of high-confidence sequence polymorphisms and its application to explain the rich phenotypic diversity among a large set of cultivated tomato accessions and its wild relatives. We observed group, species, and accession specific polymorphisms some of which can be attributed to economically important fruit and growth traits. Such information can easily be translated into array or PCR based assays to genotype genetic variants across extensive populations as well as a population of progeny. Provided that gene models from the Heinz annotated reference genome are also applicable for the other species, we observe 8% to 10% of all sequence polymorphisms to be located in the genic portion of the genome. Non-synonymous and synonymous SNPs each occur at 1% of the total number of SNPs. As a conserved estimate for wild species it would equal to about 1x $10^5$ of non-synonymous SNPs, but little is known how much of the phenotypic diversity can be attributed to this. Considering that traits like fruit colour and shape are determined by a single SNP, the total number of SNPs most likely represents a wealth of diversity that waits exploration. We are nevertheless at the beginning of elucidating their biological relevance.

*Relationships of tomatoes and wild species relatives*

The past few decades have seen the publication of several phylogenetic studies of *Solanum* species in the *Lycopersicon* section, but usage of phenotypic characters, markers and sequencing data resulted in dissimilar trees, with provisional species groupings lacking fully resolved relationships (Grandillo *et al.*, 2011). In this study we reconstructed intra and interspecies relationships for a large number of tomato accessions and related wild species, taking advantage of whole genome sequence data to maximize tree resolution. Initially our phylogenetic analysis focused on genes controlling economically important traits that have been subject to interspecific hybridization breeding. Since a subset of these genes originated from wild species, cladistics potentially may result in skewed relationships. Yet, our Maximum Likelihood consensus cladograms for the targeted genes and for the whole genome SNP sets show a comparable tree topology down to the sub sectional (species group) level, suggesting that phylogenetic relationships between fruit and growth diversification genes are not particularly biased. While, the strict consensus cladogram for the concatenated fruit and growth diversification genes displays unresolved relationships at the species level for some of the cultivated tomato and *S. peruvianum* accessions, the use of whole genome SNP data allowed increased tree resolution. Indeed, the whole genome SNP set supports the placement of taxa into separate branches with high bootstrap values for each of the accessions and wild species, including corrected placement of several previously putative typed accessions.

Ecological differences probably have resulted in dramatic genome evolutionary consequences. Moreover, there is evidence that mating system shifts have a large impact on complex multigene based traits such as floral and fruit development (Moyle, 2008), which might further account for a large intraspecies variation. Large intraspecies variation trends have been observed for *S. chilense*, which have been grouped into geographic races that can be distinguished both morphologically and genetically (reviewed in Grandillo *et al.*, 2011). Other examples involve remarkable levels of morphological and genetic diversity found in *S. peruvianum* populations (Rick, 1986; Städler *et al.*, 2005 and references therein), which might explain the distinct phylogenetic positions for *S. arcanum* LA2157 and LA2172. Here, we placed both accessions into the *Arcanum* group with northern species of the *peruvianum* complex. Accession LA2172, which is allogamous SI, appears sister to the monophyletic *S. neorickii* clade, while LA2157 is facultative SC and sister to the monophyletic *S. chmielewskii* clade. Furthermore, it is important to note that AFLP cladistics previously resulted in the grouping of *S. arcanum* LA1984 with southern *S. peruvianum* species, while the other *S. arcanum* accessions grouped with *S. huaylasense* (Spooner *et al.*, 2005). Interestingly, it has been speculated that accessions such as LA1984 could represent a 'crossing bridge' between morphologically and genetically distinct populations (Rick, 1986; Grandillo *et al.*, 2011).

*Detection of introgressions in crop accessions and genome structure*

While marker assisted introgressions focuses on the relations between traits and allelic variants, it is mostly used for the indirect selection of genetic determinants for a trait of interest and is restricted to alleles that can be diagnosed. Based on genome wide SNP data, introgressions from *S. pimpinellifolium* into chromosomes 4, 9, 11 and 12 of *S. lycopersicum* Heinz1706 were previously reported (The Tomato Genome Consortium, 2012). Following the same strategy, here the bulk of our introgression detection was based on SNP distributions divergent from the reference genome targeting introgressions not present in Heinz. Our approach shows that both location and size of introgressed segments can be inferred from the SNP distribution. Furthermore, based on the phylogenetic distance we assigned a closely related wild species *S. pimpinellifolium* as the most close donor species among the 84 accessions that we have tested. These results put a new perspective to future introgression hybridization breeding.

The success of introgressive hybridization breeding depends, among others, on the proper identification of colinear chromosome segments in donor and recipient genomes, which in turn is dependent on the consistency and completeness of their assembled genomes. The genome structure of the parental species influence crossing success and a difference in genomic colinearity has a direct effect on chromosome pairing at meiosis and hence determines the rate of alien chromatin transfer into a recipient crop. However, the proper ordering and orientation of contigs into megabase sized scaffolds depends on the availability of genetic and physical maps, which are currently lacking for the three de novo sequenced genomes. Furthermore, the N50 contig sizes for de novo assemblies of *S. arcanum*, *S. habrochaites* and *S. pennellii* do not exceed 400kb. Although the advances in next-generation sequencing technology for the use of extant germplasm resources now allow relatively fast and cheap assembly of large numbers of complex genomes, it does not yet allow a full genome reconstruction of the Solanaceae family and hence is yet of limited use for introgression breeding. The identification of compatible genomes for introgression breeding, the rearrangement phylogeny within the Solanaceae, and reconstruction of the ancestral Solanum karyotype all require additional physical mapping information on top of the genome sequence information to properly order contigs along the chromosome arms. Therefore, we believe there is room, in the near future, to pursue the integration of NGS and new technological platforms to advance the Solanum genome reconstruction.

## Experimental procedures

### *Selection of tomato accessions*

We genotyped the 7000 accessions in the EU-SOL project (https://www.eu-sol.wur.nl) on the basis of 20 traits and markers, followed by a denser genotyping of a subset of 1000 accessions using 384 SNP markers, and a final selection of 200 accessions covering the full genetic diversity of the crop. We also included a set of old cultivars that were selected on the basis of previously documented trait identifications of wild tomato relatives (reviewed in Grandillo *et al.*, 2011).

### *DNA isolation*

Young leaves were collected from the first plant of each plot (self-compatible accessions) or from the pollen acceptor (self-incompatible accessions) for DNA extraction. Approximately 100 mg frozen leaf material was grinded using the Retch Mixer Mill M300. Subsequently, genomic DNA was extracted with a standard DNA isolation protocol (Bernatzky and Tanksley, 1986), using a nuclear lysis buffer with sarkosyl. The DNA was quantified using the Qubit 2.0 Fluorometer (Invitrogen). Per accessions 1,5 – 2,0 µg DNA was used for library construction.

### *Illumina and 454 libraries sequencing and read mapping*

Shallow sequencing of 500bp inserts was carried out using Illumina HiSeq 2000 to generate a 100bp paired end library at an average of 36 fold coverage. Bases with Q < 20 were trimmed before read mapping with BWA (Li and Durbin, 2009, Li and Durbin, 2010) against *S. lycopersicum* cv. Heinz v2.40 with a maximum insert size of 750bp (50% deviation), reporting at most 30 hits and removing PCR duplicates. SAMTOOLS (Li *et al.*, 2009) was used for variant calling without skipping InDels, a minimum gap distance of 5bp, a minimum alignment quality of 20, a minimum depth of 4 and otherwise the default parameters. The same protocol was used to map the wild species to their closest *de novo* version 1.0 assembled counterpart. Contamination with *Escherichia coli*, human, insects, mouse, Phi X 174, yeast and phytoviral genomes (Adams and Antoniw, 2006) was checked with BOWTIE (Langmead *et al.,* 2009).

*De novo assembly of the three wild species genomes and Heinz*

For *de novo* sequencing of *S. arcanum* LA2157, *S. habrochaites* LYC4 and *S. pennellii* LA0716 we sequenced an overlapping paired end library with 170bp insert size, at 93.2, 76.4 and 80.8 fold coverage; re-used the 500bp insert size paired end library at 35.7, 35.6 and 28.2 fold coverage, using 100bp Illumina HiSeq 2000 reads; and a mate pair library with 2kbp insert size at 33.8, 38.0 and 31.2 fold coverage, respectively. Using 454 FLX a long mate pair library of 8kbp insert size and an extra-long mate pairs library of 20kbp insert size was sequenced at 0.55±0.10 and 0.47±0.07 fold coverage, respectively. For *S. pennellii* LA0716 we sequenced an additional short mate pair library of 3kbp insert size at 0.4 fold coverage. On average, reads produced from 454 libraries contained 35%±7% of adaptamers. *S. lycopersicum* cv Heinz 1706 used a reduced set of its original set or reads with 14.78 fold 250 PE, 17.54 fold 300 PE, 37.42 fold 500 PE, 6.25 fold 2kb MP, 6.51 fold 3kb MP, 5.94 fold 4kb MP and 6.02 fold 5kb MP in a total of 69.74 fold coverage for PE libraries, 24.73 fold coverage for MP libraries and 94.47 fold coverage overall.

The *S. pennellii* and *S. habrochaites* data were assembled with ALLPATHS-LG (assembly version 2.0) according to Gnerre *et al.* (2010) with ploidy of 2, while *S. arcanum* was assembled using CLC Genomics Workbench v7 (CLC Inc, Aarhus, Denmark) with a bubble size of 300, a minimum contig length of 200 and a word size of 64 (assembly version 1.0). Subsequently, *S. arcanum* was assembled with ALLPATHS-LG (assembly version 2.0). ALLPATHS-LG generated scaffolds were further scaffolded using the 454 FLX data in the SCARPA scaffolder (Donmez and Brudno, 2013). Subsequently, the *de novo* assembly statistics were compared to the tomato reference genome *S. lycopersicum* cv. Heinz version SL2.40 (table S2). The CLC, ALLPATHS-LG, and the ALLPATHS-LG plus SCARPA assembly is referred to as Version 1.0, 2.0 and 3.0, respectively. *S. arcanum* V1.0, *S. habrochaites* V2.0 and *S. pennellii* V2.0 were used for the mapping of the 84 accessions. Version 3.0 was used to assess genome sizes and rearrangements for all species.

*Sequence diversity and phylogenetic relationships of 84 accessions*

To assess sequence diversity in domestication syndrome genes, orthologs in 84 accessions were obtained from reciprocal best BLASTN hits of CLC assembled contigs and tomato ITAG 2.4 annotated sequences (http://solgenomics.net/gbrowse/bin/gbrowse/ITAG2.3_genomic) and aligned with CLUSTALW (Thompson *et al.*, 1994). SNPs were then called using the quality-based variant detection algorithm in CLC. Optimal substitution models for CLUSTALW-aligned gene and concatenated gene sequences were calculated in MEGA5.1 (Tamura *et al.*, 2011). Maximum Likelihood trees for each individual gene as well as concatenated gene sequences were inferred using a Neighbour-Joining initial tree (NJ) followed by Nearest-Neighbour Interchange (NNI). Phylogenies were tested using 1000 random genes separated in 100 sets of 10 genes (fig. S6). Finally, strict consensus trees for individual genes and concatenated genes were calculated using a cut-off value of 50%.

For each species we used a concatenation of all homozygous non-unique SNPs (Van Gent *et al.*, 2011) with quality above 20, which were obtained from the VCF files generated by BWA and SAMTOOLS. Multiple Nucleotide Polymorphisms (MNP) and Insertion-Deletion events (InDels) were disregarded due to their low frequency and the low aligment speed. We used ITAG v2.3 gene models with the FASTTREE 2.1.7 software (Price *et al.*, 2010) for a heuristic neighbor-joining as input to the approximately-maximum-likelihood algorithm, thus reducing the number of trees with a mix of nearest-neighbor interchange (NNIs) and subtree-prune-regraft. A Jukes-Cantor generalized time-reversible model, bio neighbor joining (BioNJ) weighted joins, 100 bootstrap resamples and gamma fitting for reported likelihood were used in the analysis.

*Annotation of SNP calls*

All VCF files from the mapped individuals were processed using SNPeff 3.4 (Cingolani *et al.*, 2012) base on ITAG 3.1 annotation and default parameters. SNPeff annotates SNP in the VCF files, based on their position and the reference annotation, with their effects and reports statistics such as rates of synonymous and non-synonymous SNPs (figs. S2 and S6), heterozygosity levels (figs. 3 and S3), the number of SNPs per 1 Mbp bins (fig. S4) and location of the SNP (fig. 4).

*Introgression estimation in S. lycopersicum*

To estimate the level of introgression in the *S. lycopersicum* species, we used the median polish procedure (Mosteller and Tukey, 1977 and Xie *et al.,* 2009) on the table of SNP count for each accession per 1Mbp bin along each chromosome (fig. S4, left panel) to remove species and bin specific effects (species or bins naturally having a higher number of SNPs). The residuals were tested using a *z*-test and bins from crop accessions with residuals significantly different from the average ($p < 0.05$) were labeled as introgressions (fig. S4, right panel). Note that in wild *S. lycopersicum* we cannot discriminate between natural variance or interspecific crossings.

*Variant browser*

JBROWSE 1.10.12 (Skinner *et al.*, 2009) was set-up to visualize the detected structural variants. The SL2.40 genome assembly and ITAG 2.31 genome annotation was loaded together with the VCF files of the 84 accessions.

*Sequence repository*

Sequence reads and associated analyses are deposited at the European Nucleotide Archive (http://www.ebi.ac.uk/ena/) under PRJEB5226 (*S. arcanum* LA2157), PRJEB5227 (*S. habrochaites* LYC4), PREB5228 (*S. pennellii* LA0716), and PRJEB5235 (reseq accessions).

## Acknowledgements

This research was supported by The Technological Top Institute Green Genetics (TTI GG), financial aid from the Dutch Ministry of Economic Affairs, Agriculture and Innovation, Centre for BioSystems Genomics (CBSG) and additional funding of industrial partners listed in the affiliates section.

**Short legends for supporting information**

**Fig. S1:** Genome wide SNP counts from 84 accessions versus reference genome of *S. lycopersicum* Heinz 1706.

**Fig. S2:** Non-synonymous (dN) and synonymous (dS) SNPs in tomato accessions and related wild species.

**Fig. S3:** Ratio between heterozygous and homozygous SNPs by mating type.

**Fig. S4:** Heatmap of introgressed regions

**Fig. S5:** Introgression in *S. lycopersicum* cv MoneyMaker

**Fig. S6:** Average and standard deviation of synonymous and non-synonymous SNPs in 100 random groups of 10 genes plus 10 syndrome genes.

**Fig. S7:** Accession specific fruit colour traits.

**Fig. S8:** Accession specific fruit shape traits.

**Fig. S9:** Strict consensus tree based on 10 fruit and growth diversification gene sequences from 84 accessions.

**Fig. S10:** Strict consensus tree based on homozygous SNPs in the exons of ten fruit-and-growth-diversification-genes sequences from 84 accessions.

**Table S1:** Selected tomato and wild species accessions.

**Table S2**: Assembly statistics.

## Figure legends

**Fig. 1:** Evaluation of genome assemblies. *De novo* assemblies for *S. lycopersicum* Heinz 1706 (purple), *S. habrochaites* LYC4 (blue), *S. pennellii* LA0716 (red) and *S. arcanum* LA2157 (orange) were generated with the ALLPATHS-LG assembler and scaffolded with SCARPA scaffolder using the 454 data. The number of contigs (x-axis) is plotted against the cumulative contig size (y-axis) when contigs are ordered by size, largest first. The gold standard assembly *S. lycopersicum* cv Heinz 1706 v2.40 is plotted in green.

**Fig. 2:** Percentage reads from *S. lycopersicum*, *S. arcanum*, *S. habrochaites* and *S. pennellii* accessions mapped against reference genomes. Species are indicated on the x-axis. Bar color codes correspond to the reference genome indicated in the legend that was used for mapping.

**Fig. 3:** Strict consensus tree based on whole genome homozygous SNPs from 84 accessions with overlaid bootstrap values obtained by Maximum-Likelihood analysis. Bars show the amount of SNPs (in millions) of the different classes of polymorphisms per accession.

**Fig. 4:** Genome wide SNP ratio for 84 accessions. SNP classes are color coded as indicated. Accession IDs and SNPs percentage are indicated on the x-axis and y-axis, respectively.

**Figure S1:** Genome wide SNP counts from 84 accessions versus reference genome of S. lycopersicum Heinz 1706. Accession names are indicated below the X-axis. Species groups are indicated by color.

**Figure S2:** Non-synonymous (dN) and synonymous (dS) SNPs in tomato accessions and related wild species. The dN and dS percentage, and dN/dS ratio relative to the total number of SNPs per accession is indicated in the left and right vertical axis respectively.

**Figure S3:** Heterozygosity level by mating type. Bars heights indicate the number of assessed species per mating type. Average ratios between homozygous and heterozygous SNPs are indicated by horizontal line while its standard deviations are represented by vertical bars.

**Figure S4:** Heatmap showing the raw SNP counts along 12 chromosomes for 54 S. lycopersicum accessions in 1Mbp bins intervals (left panel) and In the right panel, regions with SNP counts above average after median polishing using $\alpha$ = 5% for the z-score

**Figure S5:** Sequence distance graph for tomato accessions. The top graph displays the number of SNPs (right y-axis, inverted) along chromosome 6 (x-axis) which is used to calculate the sequence distance (left y-axis). Color coded lines display the level of phylogenetic distance compared to S. lycopersicum LA2463 (Moneymaker). Please note that some colored lines overlap and are merged into a black line.

**Figure S6:** Average and standard deviation of non-synonymous (dN, top panel) and synonymous (dS, middle panel) SNPs in 100 random groups of 10 genes (black vertical lines), its average (blue line) plus 10 Fruit Diversification and Plant Growth (FDPG) genes (red line). The total number of SNPs per accession is indicated by grey bars (lower panel).

**Figure S7:** Accession specific fruit colour traits. Fruit colour, SNP position and coding change in the Psy1 and B gene underlying the fruit colour phenotype are listed accordingly.

**Figure S8:** Accession specific fruit shape traits. Fruit shape, SNP position and coding change in the OVATE gene orthologs underlying fruit shape phenotype are listed accordingly.

**Figure S9:** Strict consensus tree based on 10 FDPG genes sequences from 84 accessions and S. tuberosum as an outgroup with overlaid bootstrap values obtained by Maximum-Likelihood analysis. Species names are indicated and combined with accession numbers.

**Figure S10:** Strict consensus tree based on homozygous SNPs of the exons of 10 FDPG genes sequences from 84 accessions with overlaid bootstrap values obtained by Maximum-Likelihood analysis. Species names are indicated and combined with accession numbers.

**Table S1:** List of selected accessions with names and culture collection IDs.

**Table S2:** De novo assembly statistics for S. arcanum (LA2157), S. pennellii (LA0716) and S. habrochaites (LYC4).

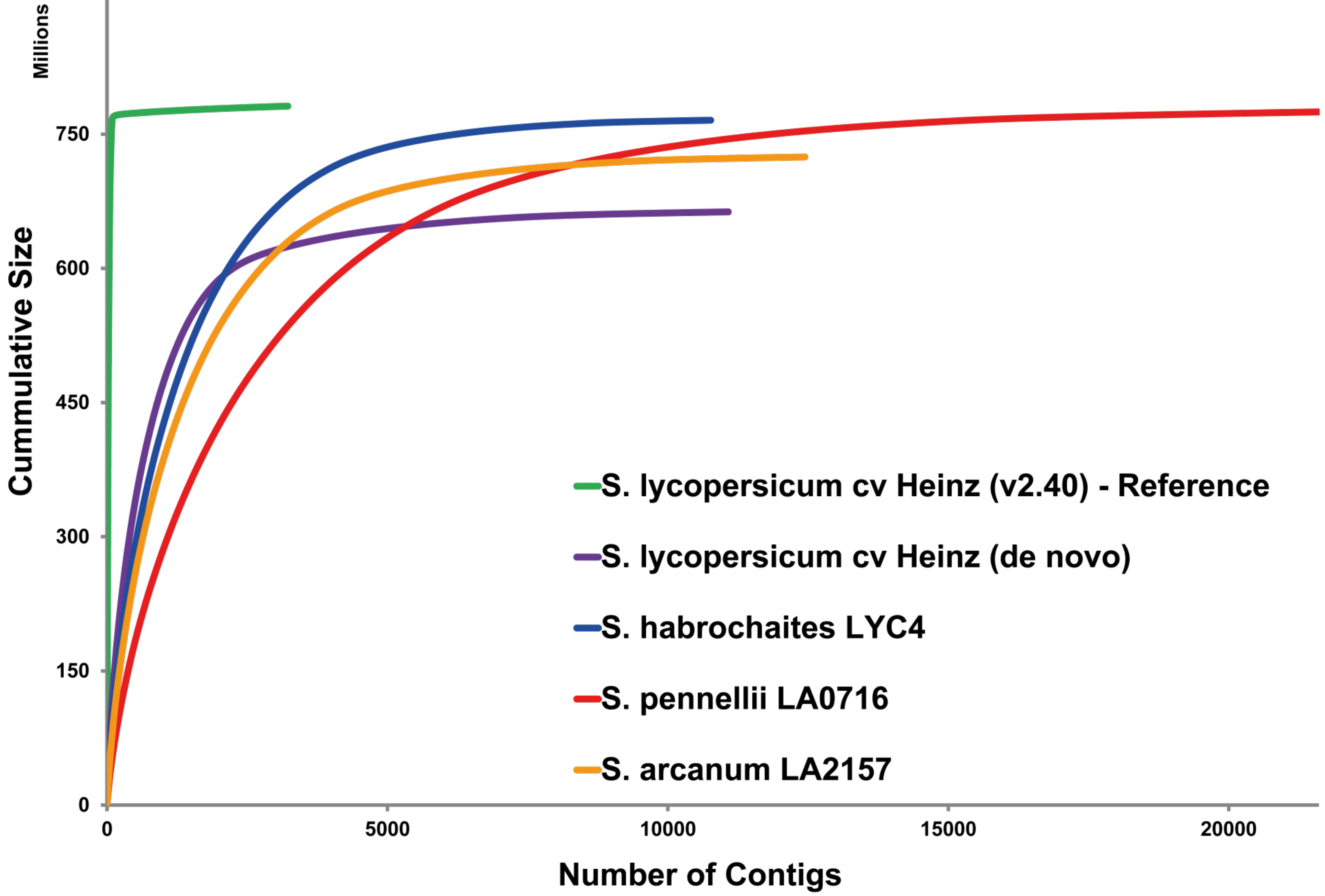

**Fig. 1:** Evaluation of genome assemblies. *De novo* assemblies for *S. lycopersicum* Heinz 1706 (purple), *S. habrochaites* LYC4 (blue), *S. pennellii* LA0716 (red) and *S. arcanum* LA2157 (orange) were generated with the ALLPATHS-LG assembler and scaffolded with SCARPA scaffolder using the 454 data. The number of contigs (x-axis) is plotted against the cumulative contig size (y-axis) when contigs are ordered by size, largest first. The gold standard assembly *S. lycopersicum* cv Heinz 1706 v2.40 is plotted in green.

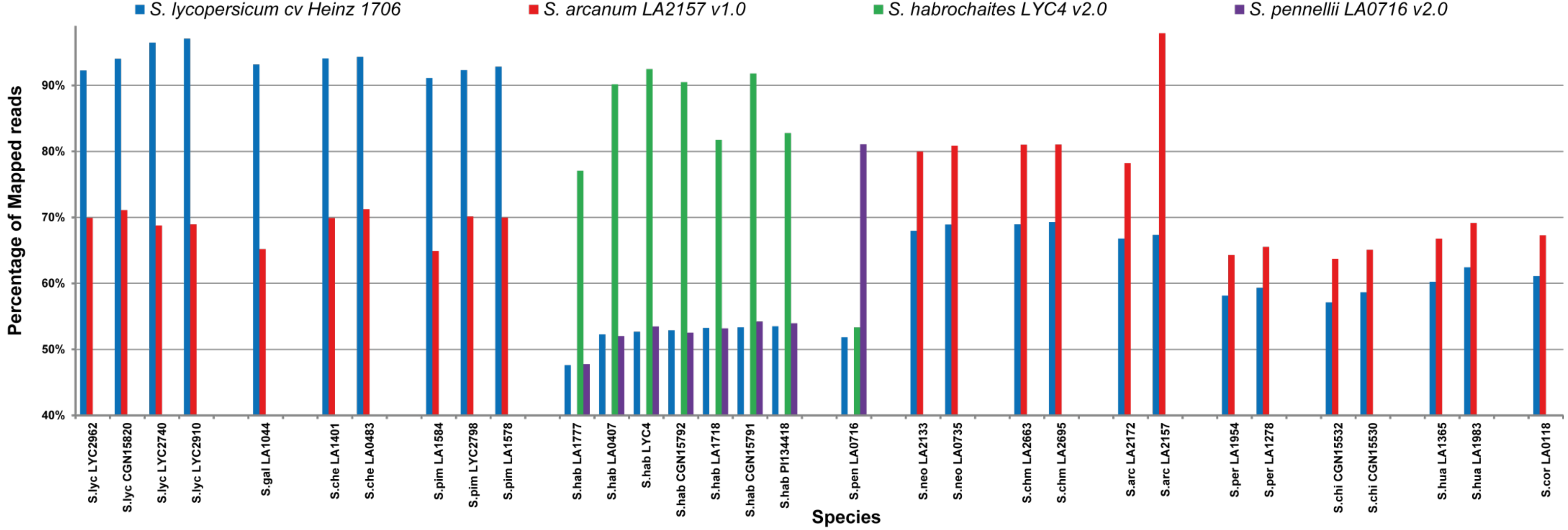

**Fig. 2:** Percentage reads from *S. lycopersicum*, *S. arcanum*, *S. habrochaites* and *S. pennellii* accessions mapped against reference genomes. Species are indicated on the x-axis. Bar color codes correspond to the reference genome indicated in the legend that was used for mapping.

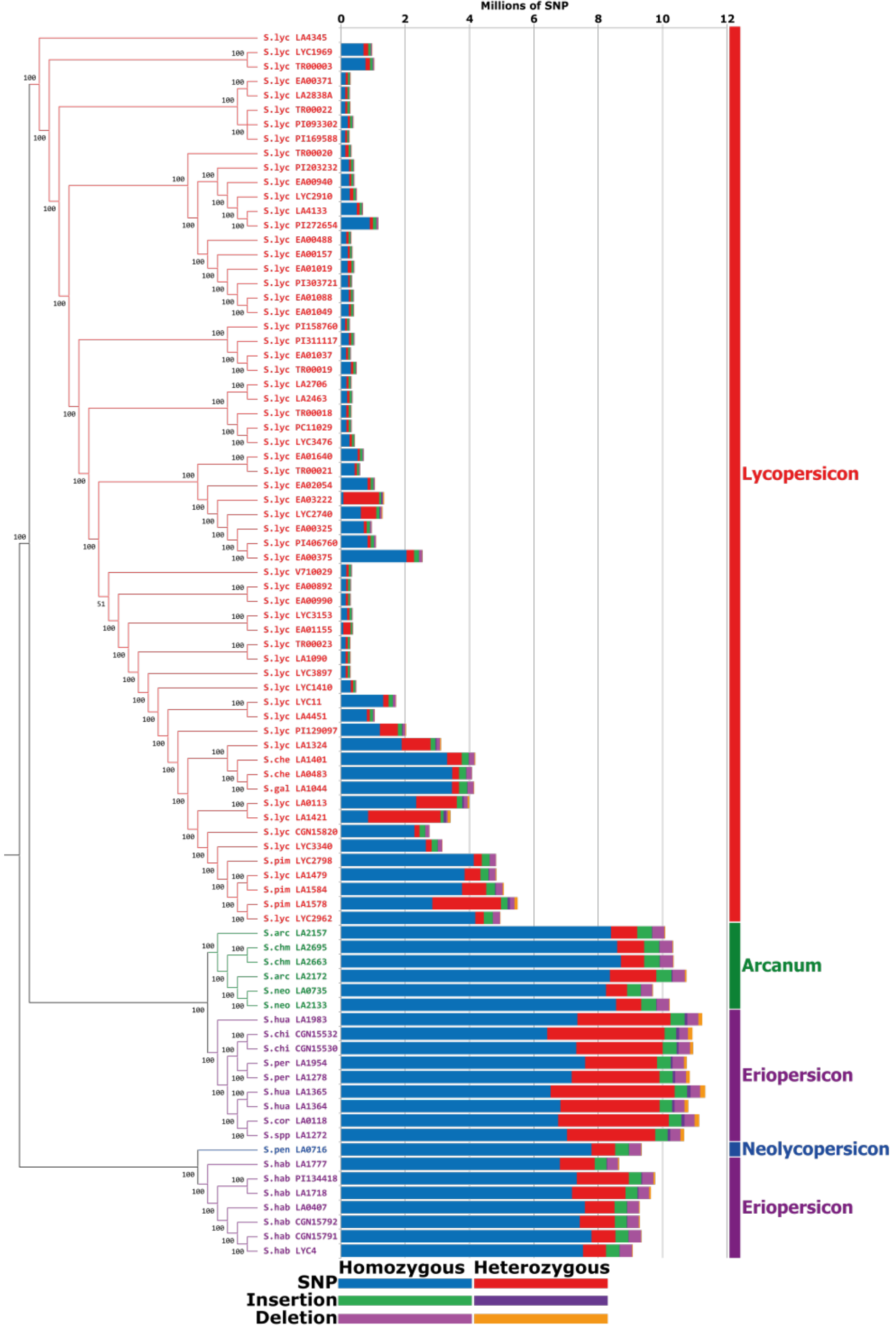

**Fig. 3:** Strict consensus tree based on whole genome homozygous SNPs from 84 accessions with overlaid bootstrap values obtained by Maximum-Likelihood analysis. Bars show the amount of SNPs (in millions) of the different classes of polymorphisms per accession

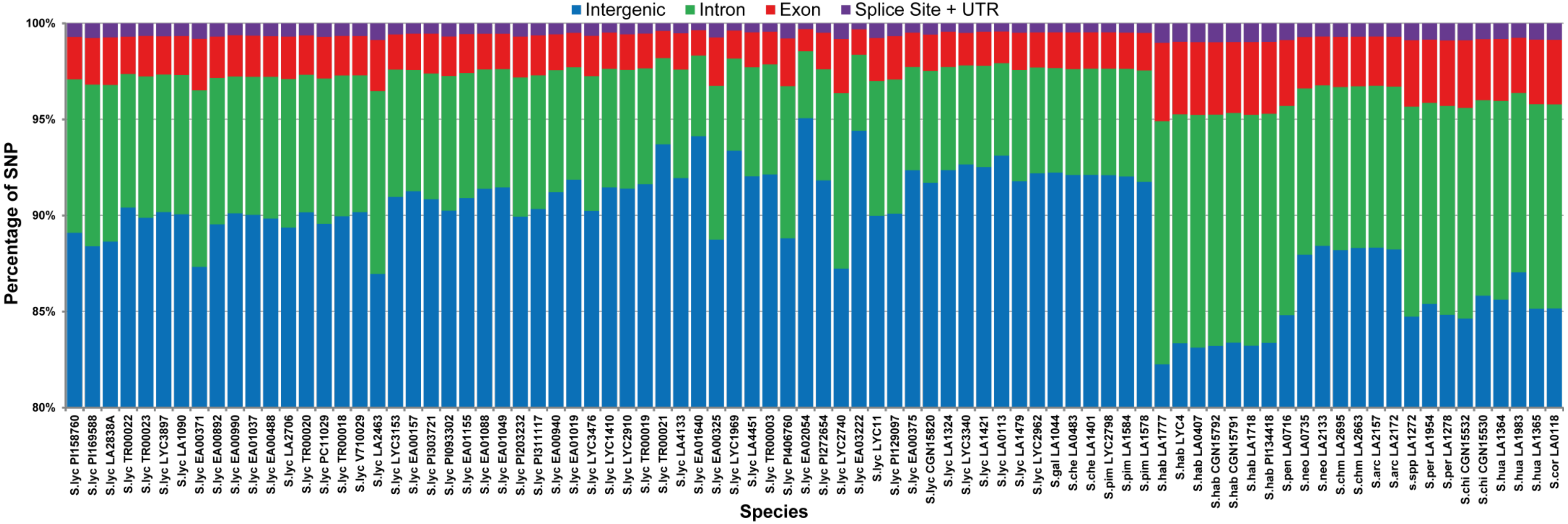

**Fig. 4:** Genome wide SNP ratio for 84 accessions. SNP classes are color coded as indicated. Accession IDs and SNPs percentage are indicated on the x-axis and y-axis, respectively

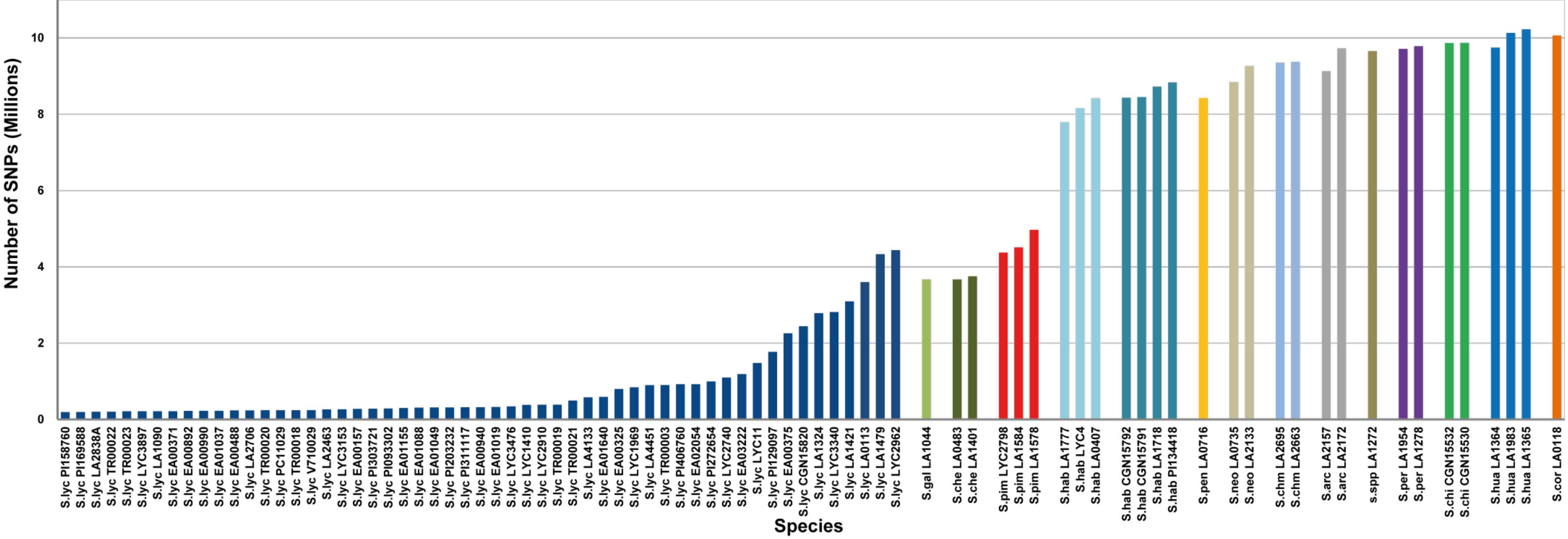

**Fig. S1:** Genome wide SNP counts from 84 accessions versus reference genome of *S. lycopersicum* Heinz 1706. Accession names are indicated below the X-axis. Species groups are indicated by color

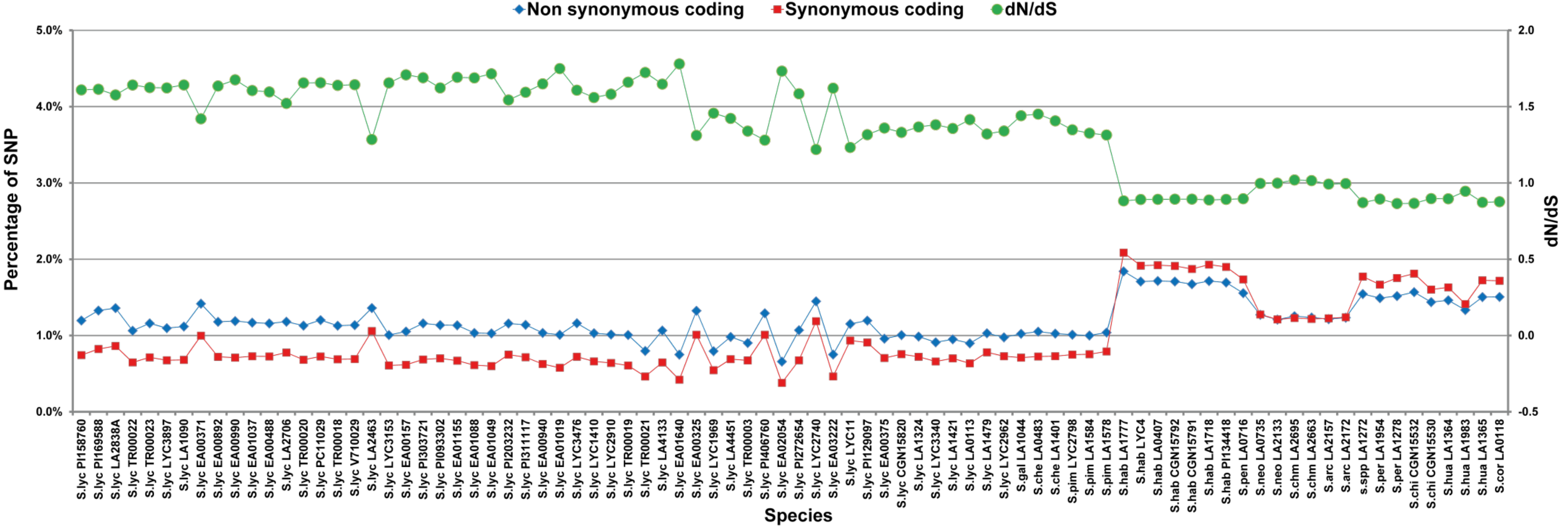

**Fig. S2:** Non-synonymous (dN) and synonymous (dS) SNPs in tomato accessions and related wild species. The dN and dS percentage, and dN/dS ratio relative to the total number of SNPs per accession is indicated in the left and right vertical axis respectively

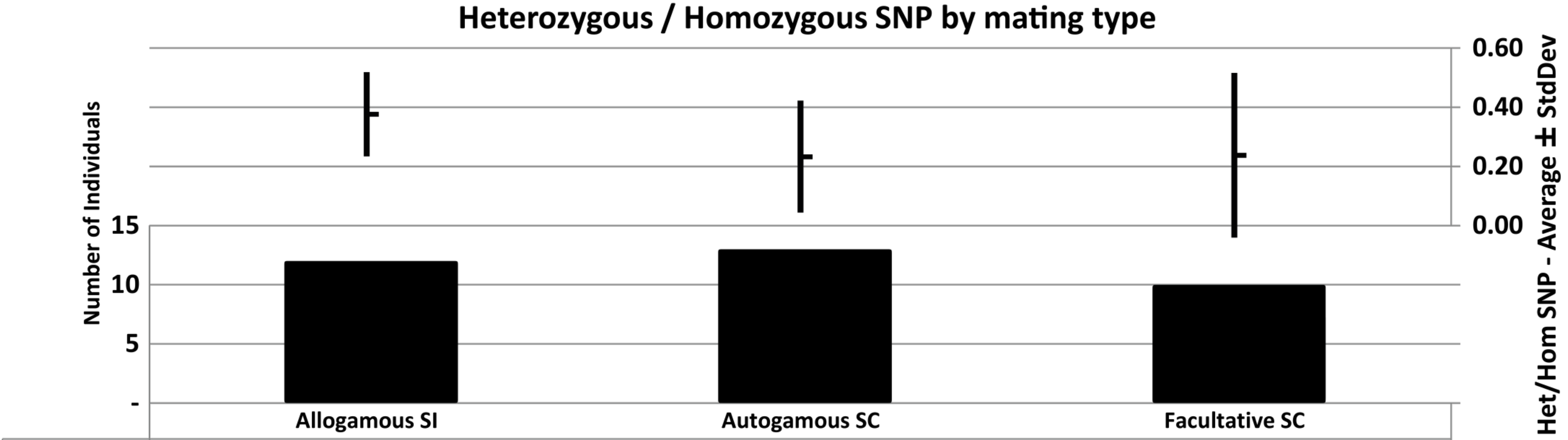

| | Allogamous SI | Autogamous SC | Facultative SC |
|---|---|---|---|
| **Number Indiv** | **12** | **13** | **10** |
| **+1δ** | **0.52** | **0.42** | **0.52** |
| **-1δ** | **0.23** | **0.04** | **-0.04** |
| **μ** | **0.38** | **0.23** | **0.24** |

**Fig. S3:** Heterozygosity level by mating type. Bars heights indicate the number of assessed species per mating type. Average ratios between homozygous and heterozygous SNPs are indicated by horizontal line while its standard deviations are represented by vertical bars

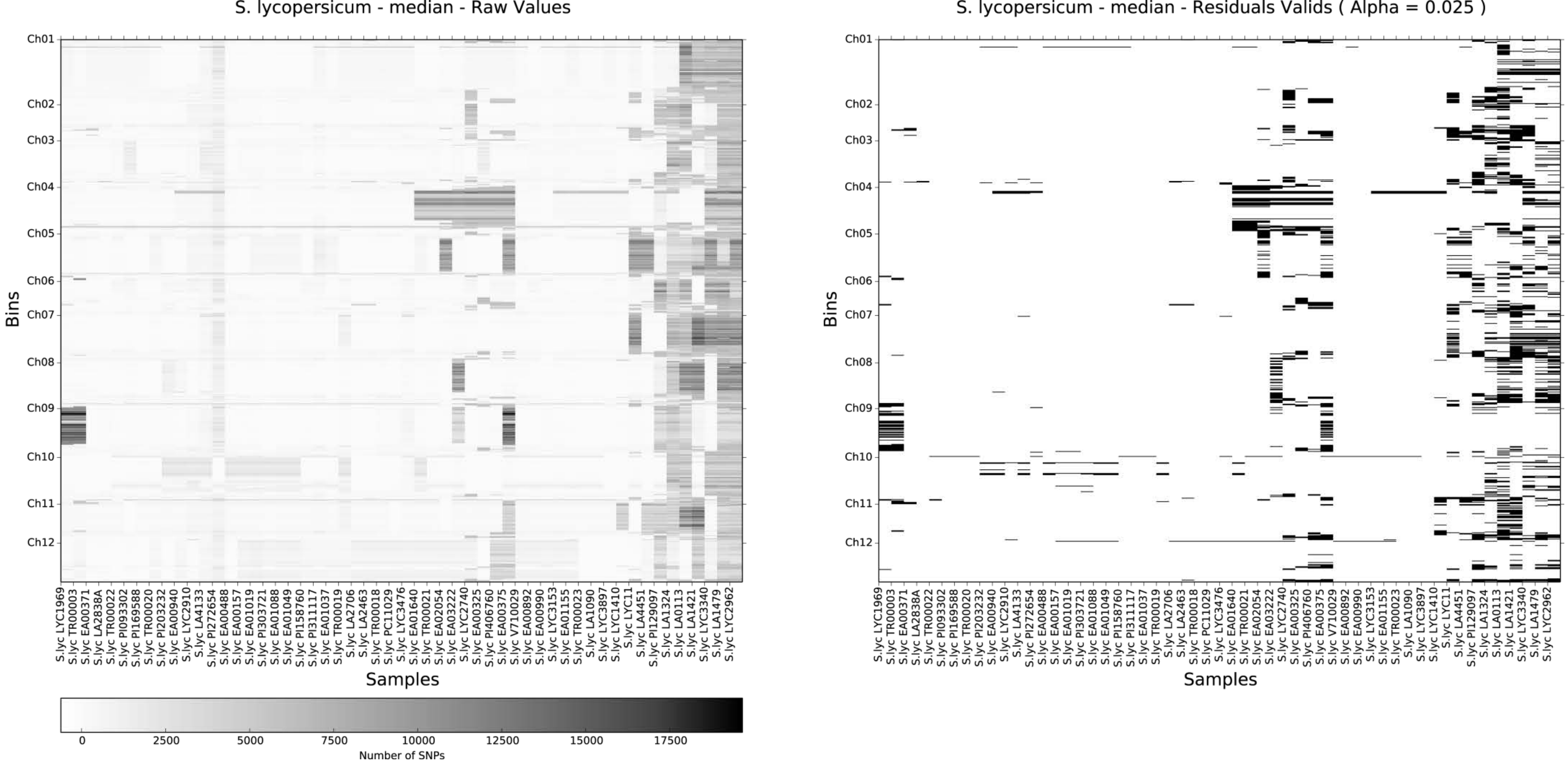

**Figure S4**: Heatmap showing the raw SNP counts along 12 chromosomes for 54 *S. lycopersicum* accessions in 1Mbp bins intervals (left panel) and In the right panel, regions with SNP counts above average after median polishing using α = 5% for the z-score

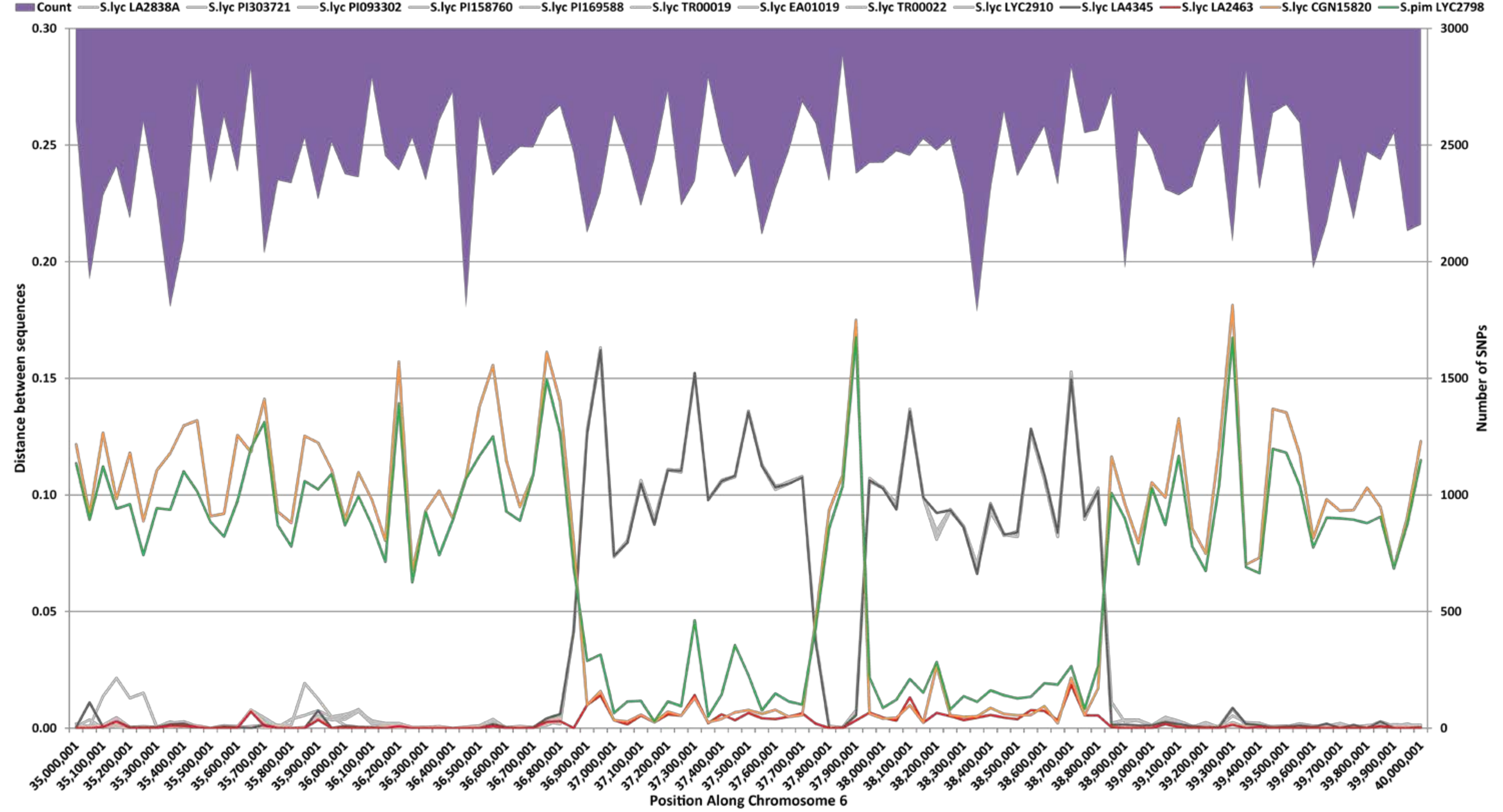

**Figure S5**: Sequence distance graph for tomato accessions. The top graph displays the number of SNPs (right y-axis, inverted) along chromosome 6 (x-axis) which is used to calculate the sequence distance (left y-axis). Color coded lines display the level of phylogenetic distance compared to *S. lycopersicum* LA2463 (Moneymaker). Please note that some colored lines overlap and are merged into a black line

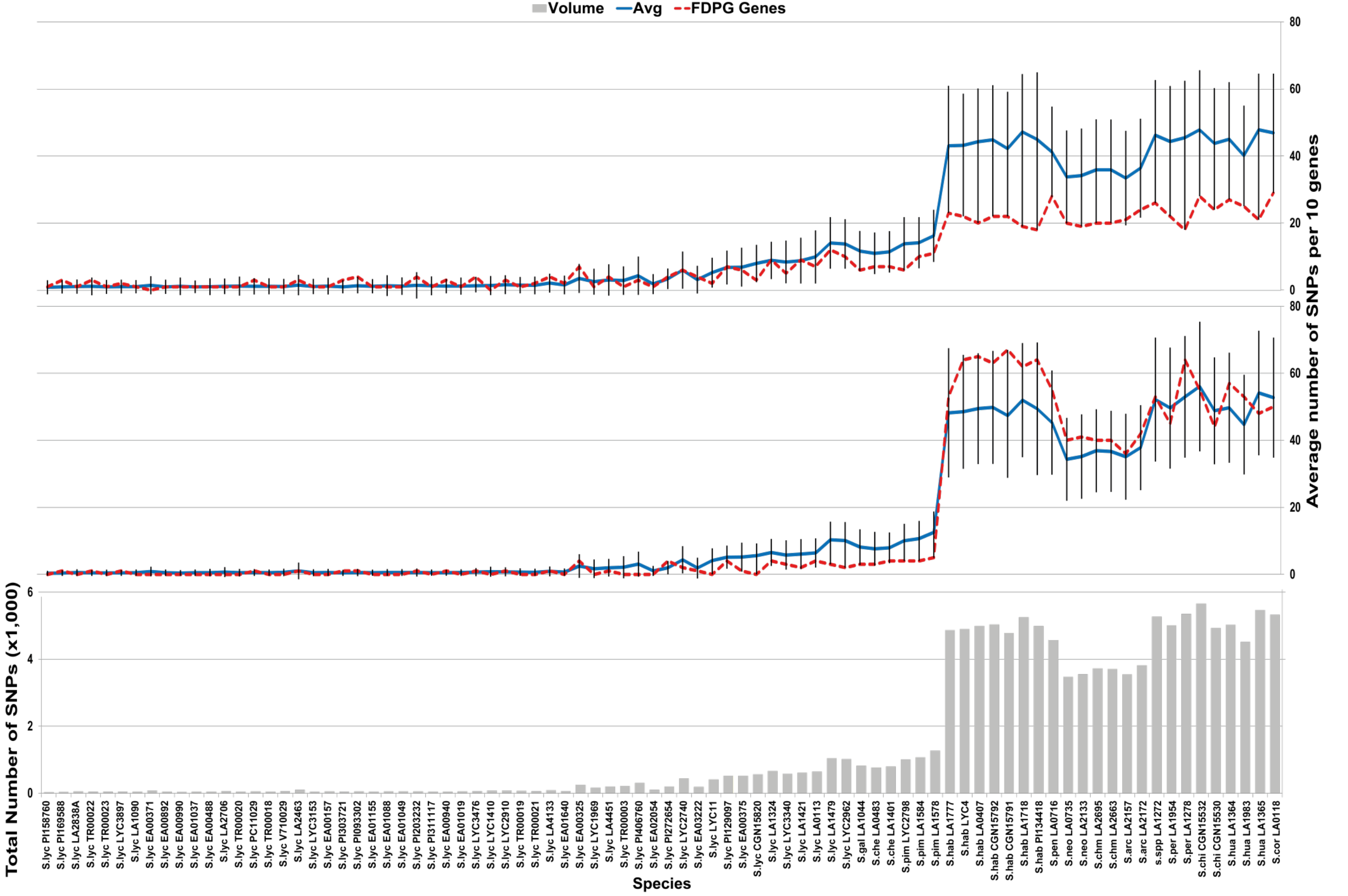

**Figure S6**: Average and standard deviation of non-synonymous (dN, top panel) and synonymous (dS, middle panel) SNPs in 100 random groups of 10 genes (black vertical lines), its average (blue line) plus 10 Fruit Diversification and Plant Growth (FDPG) genes (red line). The total number of SNPs per accession is indicated by grey bars (lower panel).

| Accession | | Color | Allele | Chr | Gene id | Mutation | Effect |
|---|---|---|---|---|---|---|---|
| 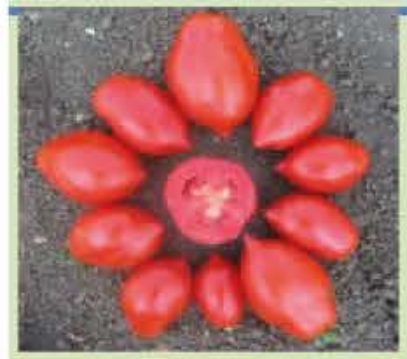 | Heinz 1706 | red | r | 3 | Psy1 | wt | Lys389 |
| 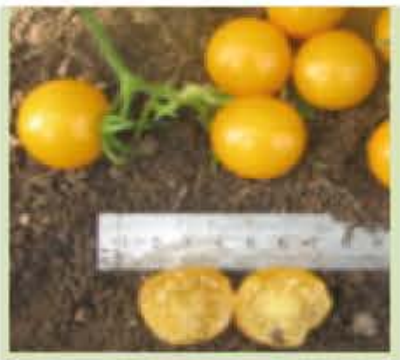 | Galina | yellow | ry | 3 | psy1 | G>del | Lys389>Ser, stop |
| 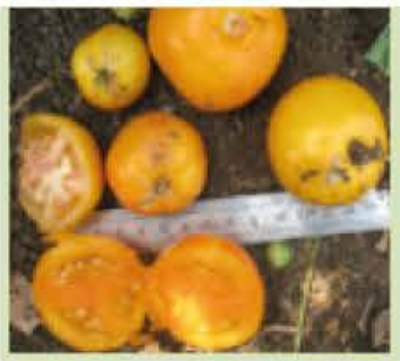 | Taxi | orange | r | 3 | Psy1 | wt | Lys389 |
| 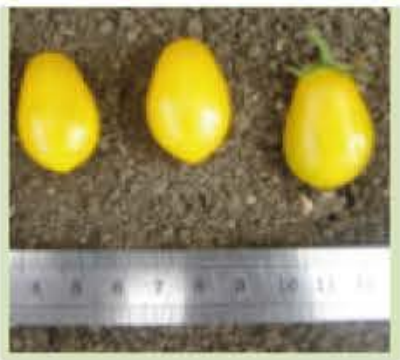 | Iidi | yellow | ry | 3 | psy1 | G>del | Lys389>Ser, stop |
| 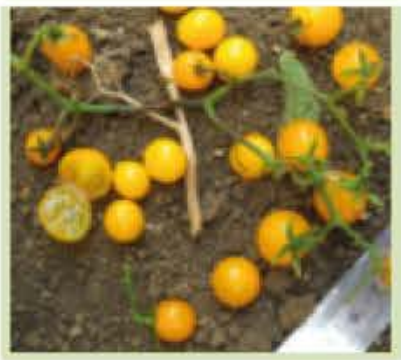 | RF17 | yellow | ry | 3 | psy1 | G>del | Lys389>Ser, stop |
| 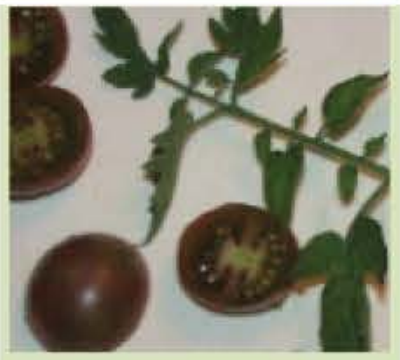 | Black Cherry | purple | ogc | 6 | b | A>del | Lys35>Asn, fs |

**Figure S7**: Accession specific fruit colour traits. Fruit colour, SNP position and coding change in the *Psy1* and *B* gene underlying the fruit colour phenotype are listed accordingly

| Accession | | Shape | Allele | Chr | Gene id | Mutation | Effect |
|---|---|---|---|---|---|---|---|
| 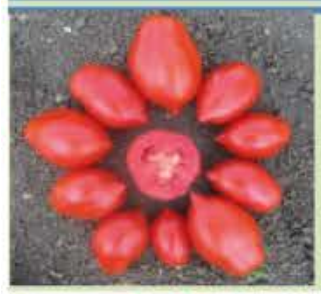 | Heinz 1706 | round | Ovate | 2 | 459212746 | wt | Glu279 |
| 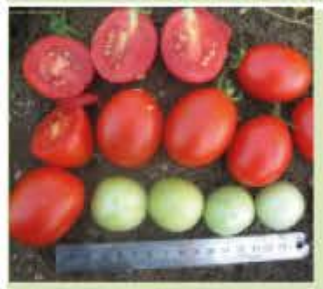 | Cross Country | pear, ovate | ovate | 2 | 459212746 | G>T | Glu279>stop |
| 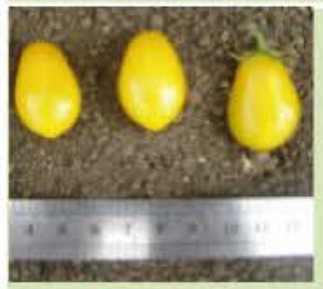 | Iidi | pear | ovate | 2 | 459212746 | G>T | Glu279>stop |
| 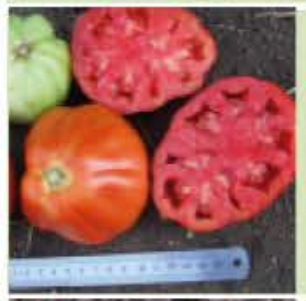 | Anto | pear, ox | ovate | 2 | 459212746 | G>T | Glu279>stop |
| 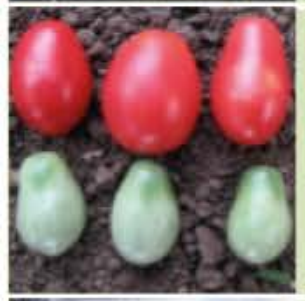 | WFR | pear | ovate | 2 | 459212746 | G>T | Glu279>stop |
| 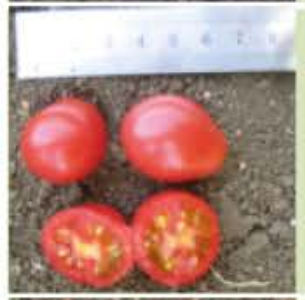 | RF36 | pear | ovate | 2 | 459212746 | G>T | Glu279>stop |
| 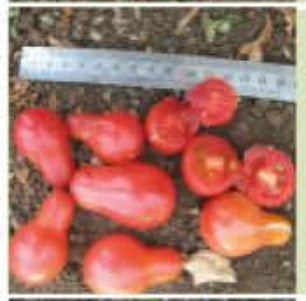 | RF43 | pear | ovate | 2 | 459212746 | G>T | Glu279>stop |
| 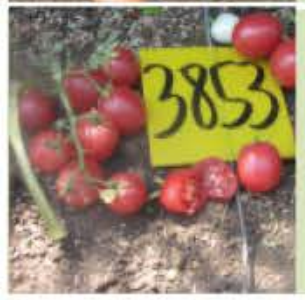  | Porter | ovate | ovate | 2 | 459212746 | G>T | Glu279>stop |

**Figure S8:** Accession specific fruit shape traits. Fruit shape, SNP position and coding change in the *OVATE* gene orthologs underlying fruit shape phenotype are listed accordingly

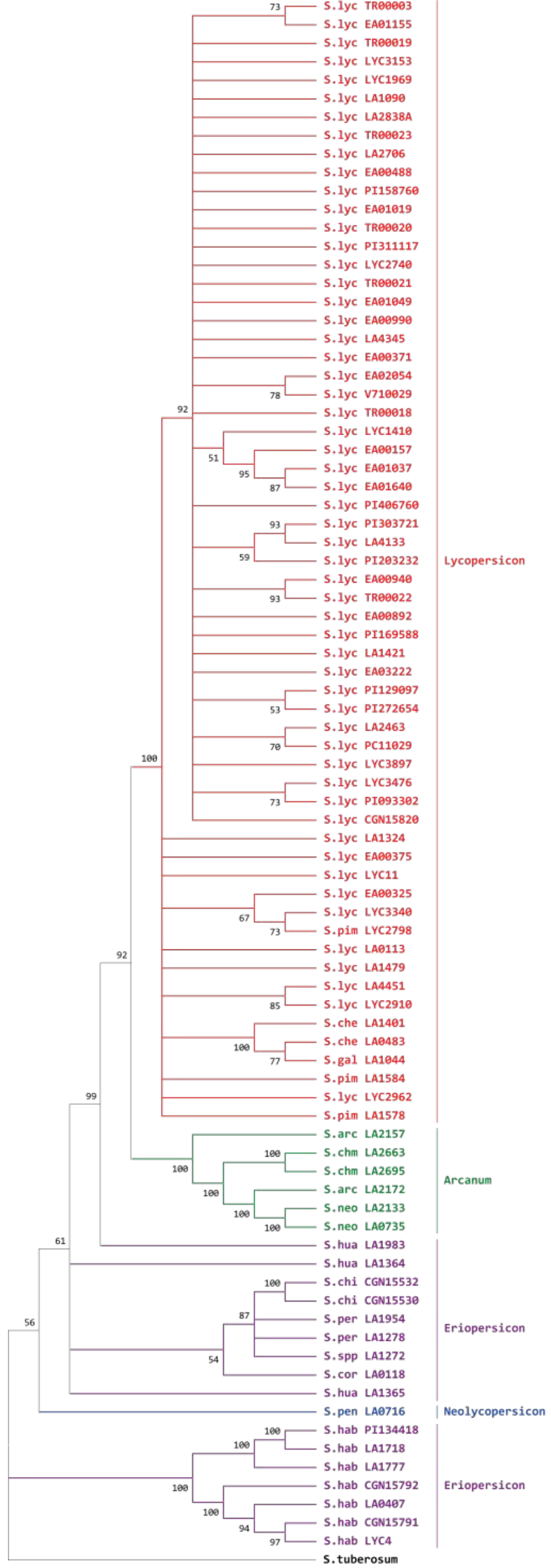

**Fig. S9:** Strict consensus tree based on 10 FDPG genes sequences from 84 accessions and *S. tuberosum* as an outgroup with overlaid bootstrap values obtained by Maximum-Likelihood analysis. Species names are indicated and combined with accession numbers

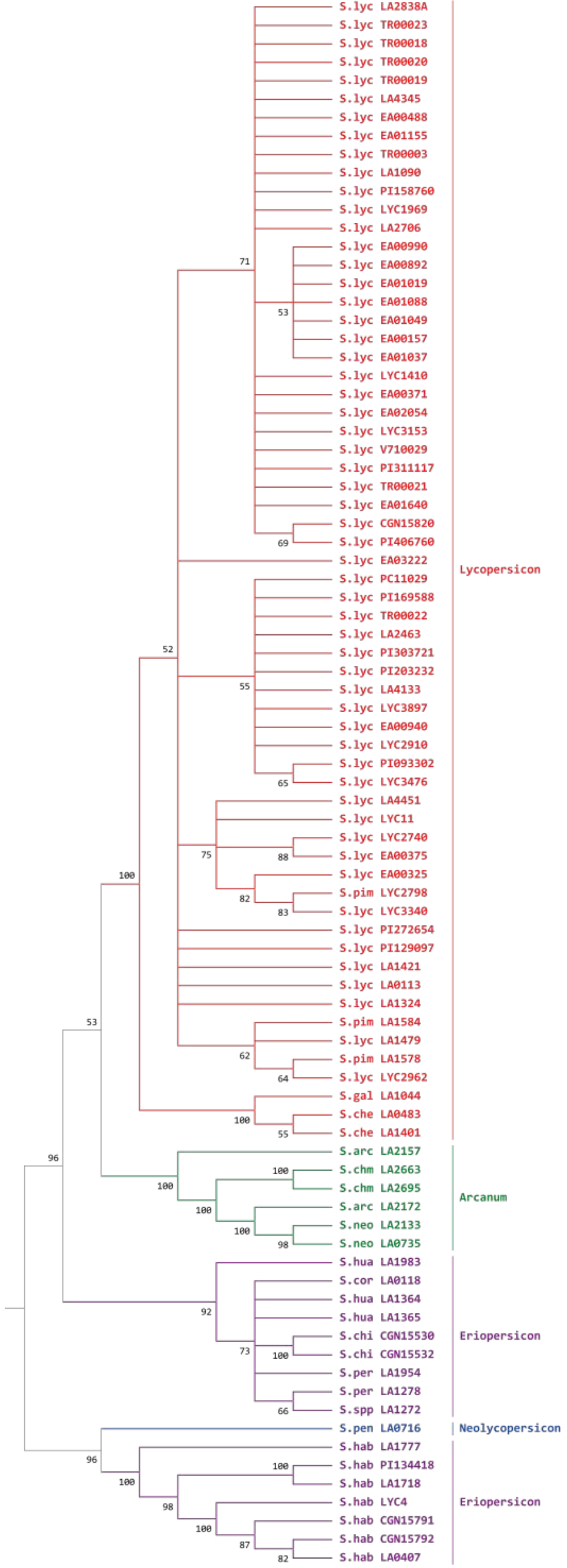

**Fig. S10:** Strict consensus tree based on homozygous SNPs of the exons of 10 FDPG genes sequences from 84 accessions with overlaid bootstrap values obtained by Maximum-Likelihood analysis. Species names are indicated and combined with accession numbers.

| Short Name | Ref | Accession name | Accession (LA/LYC/EA/PI/T/CGN/TR/V) |
|---|---|---|---|
| S.lyc LA4345 | REF | Heinz 1706 | LA4345 |
| S.lyc LA2706 | RF_001 | Moneymaker | LA2706/EA00840/EA02936/EA05097/EA10006/PI262996 |
| S.lyc LA2838A | RF_002 | Alisa Craig | LA2838A/EA01101/EA00240/EA01101 |
| S.lyc PI406760 | RF_003 | Gardeners delight | EA06086/PI406760 |
| S.lyc LA1090 | RF_004 | Rutgers | LA1090/EA00465 |
| S.lyc EA00325 | RF_005 | Galina | EA00325 |
| S.lyc EA00488 | RF_006 | Taxi | EA00488 |
| S.lyc EA00375 | RF_007 | Katinka Cherry | EA00375 |
| S.lyc EA00371 | RF_008 | John's big orange | EA00371 |
| S.lyc LA2463 | RF_011 | Allround | LA2463/LYC1365/EA02617 |
| S.lyc LYC1969 | RF_012 | Sonata | LYC1969/EA02724 |
| S.lyc LYC3897 | RF_013 | Cross Country | LYC3897/EA03701 |
| S.lyc LYC3476 | RF_014 | Iidi | LYC3476/EA03362 |
| S.lyc TR00003 | RF_015 | Momatero | TR00003 |
| S.lyc LYC11 | RF_016 | Rote Beere | LYC11/EA01965/CGN15464 |
| S.lyc LYC3340 | RF_017 | | LYC3340/EA03306/T1039 |
| S.lyc EA01155 | RF_018 | Dana | EA01155 |
| S.lyc EA01049 | RF_019 | Large Pink | EA01049 |
| S.lyc LYC3153 | RF_020 | | LYC3153/EA03221 |
| S.lyc EA03222 | RF_021 | | LYC3155/LYC2513/EA03222/T828 |
| S.lyc PI129097 | RF_022 | | PI129097/EA04710 |
| S.lyc PI272654 | RF_023 | | PI272654/EA05170 |
| S.lyc EA00990 | RF_024 | Jersey devil | EA00990 |
| S.cor LA0118 | RF_025 | S. corneliomulleri | LA0118/EA03384/T1248 |
| S.lyc EA00157 | RF_026 | Polish Joe | EA00157 |
| S.lyc EA02054 | RF_027 | Cal J Tm VF | EA02054/CGN20815*/http://www.tomaten-atlas.de/sorten/c/971-cal-j |
| S.lyc PI303721 | RF_028 | The Dutchman | EA05581/PI303721 |
| S.lyc LA4451 | RF_029 | Black Cherry | LA4451(?)/EA00027 |
| S.lyc V710029 | RF_030 | Anto | EA01835/V710029 |
| S.lyc PC11029 | RF_031 | Winter Tipe | PC11029 |
| S.lyc PI093302 | RF_032 | Chang Li | EA04243/PI93302 |
| S.lyc EA00892 | RF_033 | Belmonte | EA00892/SG16 |
| S.lyc EA01088 | RF_034 | Tiffen Mennonite | EA01088 |
| S.lyc PI203232 | RF_035 | Wheatleys Frost Resistant | EA04939/PI203232 |
| S.lyc PI311117 | RF_036 | S. lycopersicum | EA05701/PI311117 |
| S.lyc LA1324 | RF_037 | S. lycopersicum | LA1324/EA05891/PI365925 |
| S.lyc PI158760 | RF_038 | Chih Mu Tao Se | EA04828/PI158760 |
| S.lyc LA0113 | RF_039 | S. lycopersicum | LA0113/EA00526 |
| S.lyc LYC1410 | RF_040 | ES 58 Heinz | LYC1410/EA02655 |
| S.lyc PI169588 | RF_041 | S. lycopersicum Dolmalik | EA04861/PI169588 |
| S.lyc LYC2962 | RF_042 | S. lycopersicum | LYC2962/EA03107/T556 |
| S.lyc LYC2910 | RF_043 | S. lycopersicum | LYC2910/EA03058/T115 |
| S.pim LYC2798 | RF_044 | S. pimpinellifolium | LYC2798/EA02994 |
| S.lyc LYC2740 | RF_045 | S. lycopersicum | LYC2740/EA02960 |
| S.pim LA1584 | RF_046 | S. pimpinellifolium | LA1584/EA00676/PI407541 |
| S.pim LA1578 | RF_047 | S. pimpinellifolium | LA1578/EA00674 |
| S.per LA1278 | RF_049 | S. peruvianum | LA1278/PI365941/TR00005 |
| S.chm LA2663 | RF_051 | S. chmielewskii | LA2663/TR00007 |
| S.chm LA2695 | RF_052 | S. chmielewskii | LA2695/EA00759 |
| S.che LA0483 | RF_053 | S. cheesmaniae-f-minor / S. galapagense | LA0483/EA00581 |
| S.lyc CGN15820 | RF_054 | S. lycopersicum x S. cheesmaniae | CGN15820/TR00024 |
| S.che LA1401 | RF_055 | S. cheesmaniae-f-minor / S. galapagense | LA1401/EA00652/PI 365897 |
| S.neo LA2133 | RF_056 | S. neorickii | LA2133/EA00729 |
| S.neo LA0735 | RF_057 | S. neorickii | LA0735/CGN24193/TR00025 |
| S.arc LA2157 | RF_058 | S. arcanum | LA2157/TR00008 |
| S.arc LA2172 | RF_059 | S. arcanum | LA2172/TR00009 |
| S.per LA1954 | RF_060 | S. peruvianum | LA1954/EA00713 |
| S.hua LA1983 | RF_062 | S. huaylasense | LA1983/TR00010 |
| S.hua LA1365 | RF_063 | S. huaylasense | LA1365/PI 365953/TR00011 |
| S.chi CGN15532 | RF_064 | S. chilense | CGN15532/TR00012 |
| S.chi CGN15530 | RF_065 | S. chilense | CGN15530/TR00013 |
| S.hab CGN15791 | RF_066 | S. habrochaites F glabratum | PI 127827 (?)/CGN15791/TR00014 |
| S.hab PI134418 | RF_067 | S. habrochaites F glabratum | PI134418/TR00015 |
| S.hab CGN15792 | RF_068 | S. habrochaites F glabratum | CGN15792/TR00016 |
| S.hab LA1718 | RF_069 | S. habrochaites F glabratum | LA1718/EA00699/PI 390663 |
| S.hab LA1777 | RF_070 | S. habrochaites | LA1777/EA00703 |
| S.hab LA0407 | RF_071 | S. habrochaites | LA0407/EA0558/PI 251304 |
| S.hab LYC4 | RF_072 | S. habrochaites | LYC4/TR00017 |
| S.spp LA1272 | RF_073 | S. sp | LA1272/LYC1831/PI 365970/EA02701 |
| S.pen LA0716 | RF_074 | S. pennellii | LA0716/PI 246502/EA00585 |
| S.hua LA1364 | RF_075 | S. huaylasense | LA1364/TR00030 |
| S.lyc TR00018 | RF_077 | Large Red Cherry | TR00018 |
| S.lyc EA00940 | RF_078 | Porter | EA00940 |
| S.lyc TR00019 | RF_088 | Bloody Butcher | TR00019 |
| S.lyc EA01019 | RF_089 | Brandywine | EA01019 |
| S.lyc TR00020 | RF_090 | Dixie Golden Giant | TR00020 |
| S.lyc EA01037 | RF_091 | Giant Belgium | EA01037 |
| S.lyc TR00021 | RF_093 | Kentucky Beefsteak | TR00021 |
| S.lyc TR00022 | RF_094 | Marmande | TR00022/PI647486(?) http://www.solcap.msu.edu/tomato_germplasm_data.shtml |
| S.lyc TR00023 | RF_096 | Thessaloniki | TR00023 |
| S.lyc EA01640 | RF_097 | Watermelon beefsteak | EA01640 |
| S.lyc LA4133 | RF_102 | S. lycopersicum | LA4133/TR00026 |
| S.lyc LA1421 | RF_103 | S. lycopersicum | LA1421/TR00027 |
| S.gal LA1044 | RF_104 | S. galapagense | LA1044/TR00029 |
| S.lyc LA1479 | RF_105 | S. lycopersicum | LA1479/TR00028 |

| Name | N25 | I25 | N50 | I50 | N75 | I75 | N90 | I90 | Longest | Shortest | Mean | Median | Num Contigs | Total length |
|---|---|---|---|---|---|---|---|---|---|---|---|---|---|---|
| S. lycopersicum cv Heinz v2.40 | 23,291,314 | 7 | 16,467,796 | 17 | 7,023,442 | 35 | 3,041,128 | 57 | 42,121,211 | 2,000 | 242,428 | 2,847 | 3,223 | 781,345,411 |
| S. lycopersicum cv Heinz de novo V2.0 | 165,328 | 609 | 87,131 | 1,944 | 41,078 | 4,574 | 18,768 | 7,902 | 963,611 | 883 | 35,483 | 14,872 | 17,744 | 629,616,014 |
| S. lycopersicum cv Heinz de novo V3.0 | 711,921 | 154 | 373,293 | 481 | 165,009 | 1,145 | 46,832 | 2,176 | 2,560,154 | 883 | 59,866 | 6,490 | 11,077 | 663,130,306 |
| S. habrochaites de novo V2.0 | 176,864 | 680 | 97,427 | 2,111 | 49,361 | 4,745 | 23,010 | 7,950 | 990,615 | 903 | 44,066 | 20,901 | 16,708 | 736,254,084 |
| S. habrochaites de novo V3.0 | 487,032 | 257 | 253,002 | 819 | 117,458 | 1,922 | 47,819 | 3,402 | 2,330,637 | 903 | 71,129 | 12,705 | 10,763 | 765,557,122 |
| S. pennellii de novo V2.0 | 128,631 | 958 | 70,609 | 2,926 | 33,224 | 6,709 | 14,641 | 11,588 | 627,531 | 887 | 27,883 | 11,066 | 26,421 | 736,687,777 |
| S. pennellii de novo V3.0 | 235,771 | 544 | 127,741 | 1,683 | 58,513 | 3,920 | 19,193 | 7,186 | 1,470,620 | 887 | 35,862 | 7,953 | 21,606 | 774,839,444 |
| S. arcanum de novo V1.0 | 35,814 | 3,716 | 16,603 | 12,382 | 4,517 | 36,954 | 1,329 | 83,146 | 241,690 | 200 | 2,869 | 428 | 290,145 | 832,461,203 |
| S. arcanum de novo V2.0 | 159,766 | 725 | 85,931 | 2,203 | 41,599 | 5,035 | 19,217 | 8,546 | 683,045 | 905 | 36,468 | 16,124 | 18,638 | 679,689,580 |
| S. arcanum de novo V3.0 | 420,446 | 292 | 221,078 | 892 | 101,738 | 2,081 | 39,892 | 3,708 | 1,856,562 | 895 | 58,224 | 9,554 | 12,443 | 724,486,902 |